\documentclass[%
 twocolumn,
 preprint,
superscriptaddress,
 amsmath,amssymb,
 aps,prx,
 10pt
]{revtex4-2}

\usepackage{graphicx}
\usepackage{dcolumn}
\usepackage{bm}
\usepackage{svg}
\usepackage{microtype} 
\usepackage{paralist}
\usepackage[normalem]{ulem}
\usepackage[unicode=true, colorlinks=true, citecolor={blue!80!black}, urlcolor={blue!50!black}, linkcolor = {blue!80!black}]{hyperref} % add hypertext capabilities
\usepackage{array}
\usepackage{siunitx}
\usepackage{wrapfig}
\usepackage{enumitem}
\usepackage{amsmath,amssymb}
\usepackage{array}
\usepackage{soul}
\usepackage{centernot}
\usepackage[titletoc, title]{appendix}
\usepackage{titletoc}
\usepackage{lineno}
\usepackage{booktabs}
\usepackage{lineno}

\usepackage{tikz}

\begin{document}

\title{Error Semitransparent Universal Control of a Bosonic Logical Qubit}

\author{Saswata Roy}
\email{sr938@cornell.edu}
\affiliation{Department of Physics, Cornell University, Ithaca, NY, 14853, USA}

\author{Owen C. Wetherbee}
\affiliation{Department of Physics, Cornell University, Ithaca, NY, 14853, USA}
\affiliation{Department of Physics, University of California, Berkeley, USA}
 
\author{Valla Fatemi}
\email{vf82@cornell.edu}
\affiliation{School of Applied and Engineering Physics, Cornell University, Ithaca, NY, 14853, USA}

\begin{abstract} 

Bosonic codes offer hardware-efficient approaches to logical qubit construction and hosted the first demonstration of beyond-break even logical quantum memory.
However, such accomplishments were done for idling information, and realization of fault-tolerant logical operations remains a critical bottleneck for universal quantum computation in scaled systems.
Error-transparent (ET) gates offer an avenue to resolve this issue, but experimental demonstrations have been limited to phase gates.
Here, we introduce a framework based on dynamic encoding subspaces that enables simple linear drives to accomplish universal gates that are error semi-transparent (EsT) to oscillator photon loss.
With an EsT logical gate set of \{$X$, $H$, $T$\}, we observe a five-fold reduction in infidelity conditioned on photon loss, demonstrate extended active-manipulation lifetimes with quantum error correction, and construct a composite EsT non-Clifford operation using a sequence of eight gates from the set.
Our approach is compatible with methods for detectable ancilla errors, offering an approach to error-mitigated universal control of bosonic logical qubits with the standard quantum control toolkit.

\end{abstract}

\maketitle

\section{Introduction}

Realization of quantum computation faces the challenge of noise destroying encoded quantum information. 
Quantum error correction (QEC) can tackle this problem by encoding logical quantum information with redundancy to correct physical errors.
Recent experiments demonstrated implementation of QEC to protect stored logical quantum information in large arrays of physical qubits \cite{Acharya2025, Acharya2023, Krinner2022,Lacroix2025} and in single harmonic oscillators \cite{Ofek2016, Ni2023, Brock2025, Sivak2023, 2509.22191, 2510.19794, Hu2019}. 
The latter approach, in which bosonic codes take advantage of the large Hilbert space of a harmonic oscillator, has emerged as a leading platform for hardware-efficient quantum information processing \cite{4rf7-9tfx, CAI202150, MA20211789} with multiple experiments crossing the break even point \cite{Ofek2016, Ni2023, Brock2025, Sivak2023, 2509.22191, 2510.19794}.

\begin{figure*}[t]
\includegraphics[scale=1]{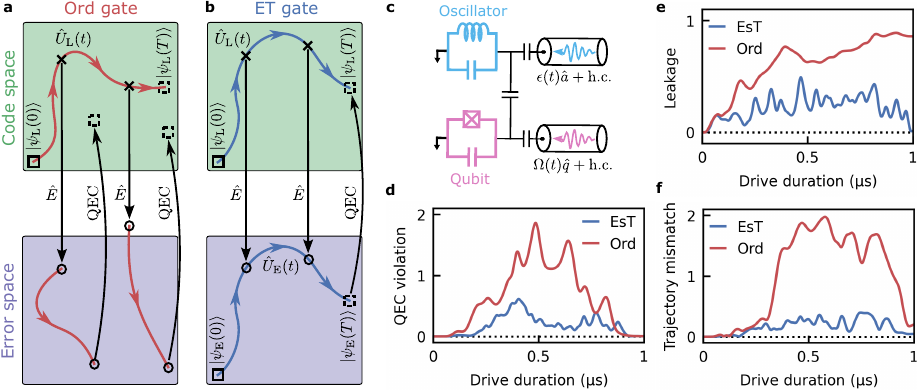}
\caption{\label{fig:fig1}\textbf{Concept and operation of ET gate:}
During a logical operation $U_L(t)$ that takes the logical code space state $|\psi_L(0)\rangle$ at time $t=0$ to $|\psi_L(T)\rangle$ at time $t=T$, an error may occur at any time $t$, taking the state from code space to an orthogonal error space.
(a) For an ordinary gate, an error at any point destroys the effect of the operation in the error space.
Importantly, the final state depends on the exact time of the error occurring, which is a probabilistic event.
(b) For an ET operation the trajectories in error and code space are identical. 
Hence an error happening at any time takes the state through a deterministic path in error space and ends at the same final state preserving the effect of the gate in error space.
Therefore a quantum error correction (QEC) process can be used to detect and correct the error at the end of the operation, making the ET operation tolerant of errors during it.
(c) An ancilla transmon qubit and an oscillator coupled to it are driven with numerically optimized pulses for realization of ET operations on an encoded logical qubit in the oscillator.
(d) QEC violation for the evolved code space basis states during a logical X operation on a binomial logical qubit encoded in the oscillator. 
Numerically optimized EsT gate has minimal violation compared to the Ord gate.
(e) Leakage out of the instantaneous error space and (f) trajectory mismatch in code and error subspaces also show smaller violation for EsT gate compared to Ord gate. 
Trajectory mismatch metric is calculated for an initial state $|0_L\rangle$.
}
\end{figure*}

Realized experiments focus on logical qubits that are protected from errors while the qubit is idling. 
Fault-tolerant quantum computing also requires fault-tolerant processing of quantum information.
In arrays of physical qubits, this is accomplished via a set of transversal gates and magic states \cite{PhysRevA.71.022316}. 
For bosonic codes, the Hilbert space is not as easily separable, so the more general formalism of error transparent (ET) gates \cite{PhysRevLett.120.050503, Vy_2013} or path-independent gates \cite{ PhysRevLett.125.110503, PhysRevResearch.4.023102} provides a path to fault-tolerant manipulation.
The fault-tolerance can be towards the errors of the oscillator and/or the ancilla.
In particular, ET gates are designed to have similar trajectories in code and error subspaces within the large Hilbert space. 
Experiments demonstrated ET \cite{ Ma2020} and path-independent \cite{Reinhold2020} phase gates in binomial logical qubits.
However, universal quantum computation requires implementation of amplitude mixing gates such as the X or H gates.
While conventional (non-ET) universal control was accomplished in Ref. \cite{Hu2019}, realization of ET universal control remains an outstanding challenge. 
Overcoming this bottleneck to achieve ET universal control would enable fault-tolerant universal computation and higher circuit depths necessary for quantum signal processing(QSP) and advanced algorithms~\cite{PRXQuantum.2.040203}.

Theoretically, the implementation of amplitude mixing gates using non-linear drives has been proposed \cite{Ma2020, Wetherbee2025mathematical}. 
These operations achieve error transparency by matching the phase and trajectory of the evolution of the code and error subspaces while staying entirely within the static code and error subspaces.
However, this approach presents significant experimental challenges due to the requirement of particularly strong nonlinear drives and extremely strict confinement to the static logical spaces, which are not currently natural or high-fidelity approaches for oscillator control~\cite{doi:10.1126/science.1259345, PhysRevX.15.021009}.

Here, we propose considering dynamic code and error subspaces as a framework to relax the static codespace requirements for error transparency.
This allows us to achieve error semi-transparent (EsT) operations using only linear drives on both the oscillator and ancilla, which is the standard optimal control `toolkit' for circuit quantum electrodynamics (cQED) \cite{Heeres2017, PhysRevA.92.040303, Eickbusch2022}.
These simpler controls also ensure fast gates, high reliability, and fewer spurious parametric processes that can complicate the dynamics.
The drawback is that, formally, exact error transparency for 
oscillator photon loss, the primary source of loss for bosonic codes, cannot be achieved because linear drive Hamiltonians cannot strictly commute with the photon loss errors on the codespace.
Nonetheless, this tradeoff can be explored for obtaining error correction gain for \textit{actively controlled} logical quantum information.

Using pulses numerically optimized for dynamic subspace error transparency to bosonic photon loss, we experimentally demonstrate universal control of a binomial logical qubit with $X$, $H$ and $T$ gates.
We compare the performance with that of ordinary, non-ET gates (henceforth named Ord gates), finding that EsT gates enhance performance in presence of photon loss errors occurring during gate operations.
We observe an average factor of five reduction in logical infidelity for the actively controlled logical states conditioned on a photon loss event during the gate.
Commensurately, the EsT gates perform significantly better than the Ord gate after applying error correction following the operation.
Further, we show that distinct gates from the set are compatible for digital construction of arbitrary unitary operations \cite{10.5555/2011679.2011685, NielsenChuang2010} while preserving the EsT property.  

\section{Theoretical concept}
We begin by defining ideal ET operations. 
As described in Ref.~\cite{Ma2020,Vy_2013,PhysRevLett.120.050503}, a logical unitary $U$ is ET if it commutes with the error set $\{E_j\}$ throughout the operation, such that $U(T,t)E_jU(t,0)|\psi_L(0)\rangle = e^{i\phi_j(t)}E_jU(T,0)|\psi_L(0)\rangle$, $\forall j, t, |\psi_L(0)\rangle$, where $U(t_2, t_1)$ describes the unitary operation from time $t_1$ to $t_2$, $T$ is the final time such that $U  = U(T,0)$, $\phi_j(t)$ is a global phase, and $|\psi_L(0)\rangle$ denotes a state in the code space at time $t = 0$.  
If the code and error spaces are static, meaning $U(t, 0)$ remains a logical unitary for all $t$, this ET condition can be satisfied by engineering the Hamiltonian to act identically within the code and error spaces: 
$P_j^\dagger H(t) P_j = P_C^\dagger H(t) P_C + d_j(t) P_C, \forall j, t$,
where, $P_j \propto E_j P_C $ is the projector from the code space to the error space corresponding to the error $E_j$, $P_C$ is the codespace projector and $d_j(t)$ is a complex number.

To relax the requirement of static code and error spaces, we propose defining an instantaneous codespace
\begin{equation}
    \mathcal{C}(t) = \{ U(t,0) | \psi_L(0) \rangle \mid | \psi_L(0) \rangle \in \mathcal{C}(0) \}~,
\end{equation} 
with instantaneous projectors $P_C(t) = U(t,0) P_C U^{\dagger}(t,0)$.
We can define a corresponding instantaneous error space ($\mathcal{E}_j(t)$) as 
\begin{equation}
    \mathcal{E}_j(t) = \textrm{span}\{ E_j |\psi_L(t)\rangle \mid | \psi_L(t) \rangle \in \mathcal{C}(t)\}~,
\end{equation} 
with projector $P_{\mathcal{E}_j}(t) \propto E_j P_C(t) E_j^{\dagger}$.
Then we need to enforce three conditions on our engineered  dynamics for error transparency.

First, the instantaneous code space must remain valid and correctable at all times, satisfying the QEC conditions (Knill-Laflamme conditions) throughout the trajectory. 
For a general error set $\{E_k\}$, this requires that the projection of all error products into the instantaneous code space is proportional to the logical identity:
\begin{equation}
    P_C(t) E_i^\dagger E_j P_C(t) = c_{ij}(t) P_C(t)~, \quad \forall i,j,
    \label{eqn:exact_qec}
\end{equation}
where $c_{ij}(t)$ are complex coefficients. 

Second, the evolved error space under the application of the unitary should coincide with the instantaneous error space defined above. 
This can be written as 
\begin{equation}
    U(t,0)\mathcal{E}_j(0) = \mathcal{E}_j(t)~, \quad \forall j,t, 
\end{equation}

Third, the Hamiltonian must drive identical dynamics within these instantaneous code and error subspaces:
\begin{equation}
    P_j^\dagger(t) H(t) P_j(t) = P_C^\dagger(t) H(t) P_C(t) + d_j(t) P_C(t)~, \forall j, t, 
    \label{eqn:ET_cond}
\end{equation}
where $d_j(t)$ is a time-dependent complex number.
Fulfilling these three conditions in the dynamic subspaces, the ET property can in principle be maintained throughout the operation. 

To quantify the deviation from ideal ET in our dynamic subspaces, we introduce three time-dependent metrics: QEC violation, instantaneous leakage and trajectory mismatch. Here we will define them analytically, and later we will assess the measures for the gates derived for our particular setup.

To evaluate deviations from ideal instantaneous correctability, we define the QEC violation metric, $\Delta_{\text{QEC}}(t)$~\cite{PhysRevA.97.032346} (see App.~\ref{app:lin_drive_limitation} for details).
Assuming a two-dimensional logical subspace, we define the $2 \times 2$ projected error matrices as $M_{ik}(t) = P_C(t) E_i^\dagger E_k P_C(t)$.
From this matrix $M_{ik}(t)$, the QEC violation metric isolates and sums the magnitude of the uncorrectable Pauli components (logical bit, phase, and joint bit-phase errors) generated across all pairs of errors $\{ E_i, E_k \}$:
\begin{equation}
    \Delta_{\text{QEC}}(t) = \sum_{i,k} \left\| M_{ik}(t) - \frac{1}{2}\text{Tr}\left[M_{ik}(t)\right] P_C(t) \right\|_2~, 
    \label{eqn:Delta_qec}
\end{equation}
where $\| \cdot \|_2$ denotes the 2-norm when applied to matrices, and the standard Euclidean norm when applied to vectors.

The instantaneous leakage, $L_{\mathcal{E}_j}(t)$, measures the weight in states outside the target error manifold associated with a specific error $E_{j}$. 
It is defined as the probability that the state, having undergone an error $E_j$, drifts outside the instantaneous error subspace $\mathcal{E}_j(t)$:
\begin{equation}
    L_{\mathcal{E}_j}(t) = 1 - \text{Tr}\left[ P_{\mathcal{E}_j}(t) \rho(t) \right]~,
    \label{eqn:leakage_t}
\end{equation}
where $\rho(t)$ is the density matrix at time $t$, when initialized as the maximally mixed state in the error space $\mathcal{E}_j(0)$ and evolved unitarily under the driven Hamiltonian without losses.

To assess the preservation of coherence within the subspace, we define the state-dependent trajectory mismatch, $\eta_{\mathcal{E}_j, {\psi}}(t)$. 
This metric quantifies the unitary rotation error by calculating the Euclidean distance between the Bloch vectors of the code state and the projected error state in the instantaneous subspaces:
\begin{equation}
    \eta_{\mathcal{E}_j, {\psi}}(t) = \left\| \vec{r}_C(t) - \vec{r}_{E_j}(t) \right\|_2~,
    \label{eqn:traj_mismatch}
\end{equation}
where the Bloch vectors are given by $\vec{r}_C(t) = \text{Tr}[\vec{\sigma}_C(t)\rho_{|\psi\rangle}(t)]$ and $\vec{r}_{E_j}(t) = \text{Tr}[\vec{\sigma}_{E_{j}}(t) {\tilde\rho}_{|\psi_{E_{j}}\rangle}(t)]$. 
Here, $\vec{\sigma}_C(t)$ and $\vec{\sigma}_{E_j}(t) \propto E_j \vec{\sigma}_C(t) E_j^{\dagger}$ are vectors of the logical Pauli operators in the instantaneous code and error subspaces.
The density matrices $\rho_{|\psi\rangle}(0)$ and $\rho_{|\psi_{E_{j}}\rangle}(0)$ are initialized as a pure logical state $|\psi\rangle  \in \mathcal{C}(0)$ and the corresponding error space state $|\psi_{E_j}\rangle \propto E_j|\psi\rangle$, respectively.
After the unitary evolution by the system Hamiltonian and Schrodinger equation, the evolved states are projected onto the instantaneous code and error spaces, respectively, and normalized to obtain $\tilde\rho_{|\psi\rangle}$ and ${\tilde\rho}_{|\psi_{E_{j}}\rangle}(t)$ (note that $\tilde\rho_{|\psi\rangle}(t) = \rho_{|\psi\rangle}(t)$ by definition of instantaneous code space).

These three metrics quantify deviations from error-transparent evolution, with optimal performance achieved as each approaches zero.
A vanishing QEC violation, $\Delta_{\text{QEC}}(t) = 0$, ensures the evolving code words remain correctable throughout the operation. 
A vanishing leakage, $L_{\mathcal{E}_j}(t) = 0$ implies the evolving error space and instantaneous error space coincide.
Finally, a vanishing trajectory mismatch, $\eta_{\mathcal{E}_j, \psi}(t) = 0$ for all initial states $|\psi\rangle$, indicates that the instantaneous error subspace evolution is identical to the code subspace evolution throughout the gate operation.
These three metrics being simultaneously satisfied at all times ensures the evolution satisfies the condition for dynamic error transparency.

\section{Application to a Binomial Code}

We now turn to applying this concept to bosonic codes, specifically the binomial `kitten' code.
To realize such EsT operations leveraging a dynamic subspace framework, we encode a logical qubit in a superconducting cavity (harmonic oscillator), dispersively coupled to a transmon qubit ancilla \cite{PhysRevB.94.014506} as shown in Fig.~\ref{fig:fig1}c. 
The system Hamiltonian at low order is given by:
\begin{equation}
    H/ \hbar = \omega_a \hat{a}^\dagger \hat{a} + \omega_q \hat{q}^\dagger \hat{q} + \chi \hat{a}^\dagger \hat{a} \hat{q}^\dagger \hat{q} + \frac{K_a}{2} \hat{a}^{\dagger2} \hat{a}^2 + 
    \frac{K_q}{2} \hat{q}^{\dagger2} \hat{q}^2~,
\end{equation}
where $\hat{a}$ and $\hat{q}$ are the annihilation operators for the cavity and transmon respectively, $\chi/2\pi =$ \SI{-3.66}{\mega\hertz} is the dispersive coupling of the transmon and cavity, $K_a/2\pi = $ \SI{-22}{\kilo\hertz} is the self-Kerr of the cavity, $K_q/2\pi = $ \SI{-180}{\mega\hertz} is the self-Kerr of the qubit (App.~\ref{app:expt_setup} describes details of experimental package).
The cavity has $T_1 = $ \SI{180}{\micro\second} and $T_2 = $\SI{290}{\micro\second} whereas the ancilla transmon has $T_1 = $ \SI{70}{\micro\second} and $T_2 = $\SI{30}{\micro\second}.
This is the same package as used in Roy, \textit{et al}~\cite{PhysRevX.15.021009}.

In the cavity, we encode the binomial `kitten' code \cite{PhysRevX.6.031006, PhysRevX.10.011058}, which allows correction of single photon loss errors in the cavity. 
The codespace for this binomial code is defined on the even parity subspace as $\mathcal{C} = \textrm{span}\{ |0_L\rangle, |1_L\rangle\}$, where $|0_L\rangle = \frac{|0\rangle + |4\rangle }{\sqrt{2}}, |1_L\rangle = |2\rangle$ are the two logical codewords. 
When a single photon loss error happens in the cavity, the logical state jumps into the orthogonal odd parity error space $\mathcal{E} = \textrm{span}\{ |0_E\rangle, |1_E\rangle\}$, where $|0_E\rangle = |3\rangle, |1_E\rangle = |1\rangle$ are the two error words. 

\begin{figure}[h]
\includegraphics[scale = 1.05]{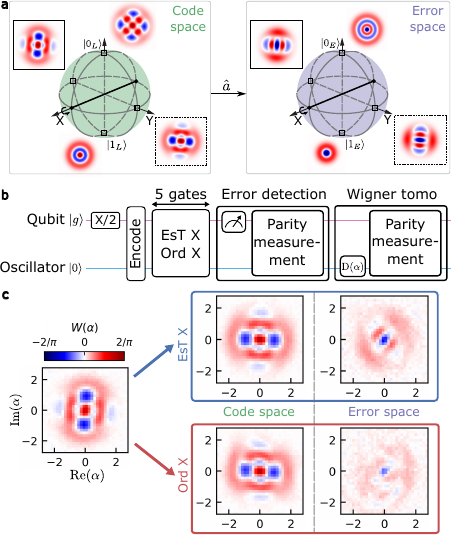}
\caption{\label{fig:fig2}\textbf{Wigner visualization of error transparency:}
(a) Logical Bloch sphere of code space (left) and error space (right) corresponding to single photon loss errors in the oscillator for binomial `kitten' code. 
The cardinal points and their theoretical Wigner functions are marked on the Bloch sphere. 
Starting from the state $\left|-Y_L\right\rangle  =(|0_L \rangle - i|1_L\rangle)/\sqrt{2}$ (Wigner function in solid square box), a logical X operation is performed along the X axis (marked black) to end in the state $|+Y_L\rangle  =(|0_L \rangle + i|1_L\rangle)/\sqrt{2}$ (Wigner function in dashed square box).
(b) Experimental pulse sequence for characterizing EsT gates. 
After the gate operation, we read out the qubit state and measure the parity of the oscillator state to detect errors and perform Wigner tomography.
Based on this parity measurement, we post-process the Wigner functions of states in code and error space. 
(c) Experimentally measured Wigner functions of binomial code cardinal point $\left|-Y_L\right\rangle  =(|0_L \rangle - i|1_L\rangle)/\sqrt{2}$ and the state after applying five EsT or Ord X gates are shown.
Left side plots of code space Wigner function is conditioned on detection of no error ($\sim 90\%$ of shots) and shows similar performance for EsT and Ord gate. 
Right side plot of error space Wigner function is conditioned on detection of single photon loss error ($\sim 10\%$) and shows EsT gate significantly preserves the phase coherence and state amplitudes of the target state compared to Ord gate.
Error space Wigners show slightly different frame rotation than codespace because of qubit being in excited state for longer during the measurement.
}
\end{figure}

\begin{figure*}[t]
\includegraphics[scale=1]{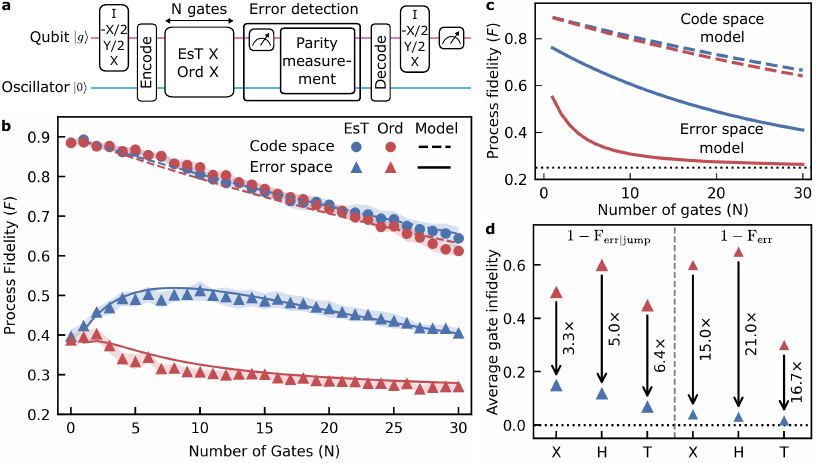}
\caption{\label{fig:fig3}\textbf{Code and error space process tomography:}
(a) Experimental pulse sequence to perform code and error space process tomography. 
Following the gate operations, we perform qubit readout and oscillator parity measurement for error detection based on which we post-select our states into code and error space. 
Separate decode operations are used that respectively maps the code and error space states to the ancilla qubit state for performing process tomography.
(b) Experimentally measured code (dots) and error space (triangles) process fidelities for EsT and Ord X gate with number of applied gates are shown (see App. C for details).
Shaded region around the data points signify 1 standard deviation of measured fidelities in repeated experiments.
Dashed lines for code space and solid lines for error space show modeled gate performance with number of gates.
In codespace, both gates perform similarly while in the error space EsT gate performs significantly better than the Ord gate.
The horizontal dashed line at process fidelity of 0.25 indicates a completely incoherent process.
(c) We plot the modeled code and error space process fidelities without non-idealities of the error detection process.
This shows a clear difference in error space process fidelity of EsT and Ord gates. 
(d) We summarize the infidelities for X, H and T gates extracted from our model for (left side of plot) error during the operation that takes a state from code space to error space ($\mathrm{F}_{\mathrm{err|jump}}$) and (right side of plot) operations in the the error space ($\mathrm{F}_{\mathrm{err}}$).
Factor of reduction in infidelity is marked by the arrow for all the cases.  
}
\end{figure*}

The flexibility of dynamic subspaces allows for the use of simpler linear drives that traverse trajectories outside this static code and error spaces to achieve fast, non-adiabatic operations while optimizing for ET (Fig. 1b) through the gate.
We consider the drive Hamiltonian
\begin{equation}
\label{eqn:drive_ham}
H_{\text{drive}}(t)/\hbar = \epsilon(t)\hat{a} + \Omega(t)\hat{q} + \text{h.c.}    
\end{equation}
However, evolutions of the codespace under linear drive on the oscillator does not generically commute with photon loss operator $\hat{a}$, the primary loss channel for the binomial code $\mathcal{C}$, on the codespace. 
Consequently, exact ET is not expected to be achieved in this setup, but trading exactness for a simpler and more scalable experimental setup may offer practical advantage and higher experimental fidelities by improving reliability and reducing spurious error channels. 
We refer to operations that deviate from exact error transparency as error semi-transparent (EsT) \footnote{The terminology draws an optical analogy: because the dynamical evolutions within the code and error subspaces are not strictly identical, the operation acts as a semi-transparent one rather than a transparent one.}.

We construct the target EsT unitary by numerically engineering the time-dependent drive Hamiltonian ($H_{\text{drive}}(t)$).
We optimize complex drive amplitudes simultaneously applied on the cavity ($\epsilon(t)$) and the ancilla qubit ($\Omega(t)$) (Fig.~\ref{fig:fig1}c), using GRAPE \cite{KHANEJA2005296, Heeres2017, PhysRevA.92.040303} for accomplishing logical $X$, $H$, and $T$ gates on the binomial code.
The Ord gates are optimized with the single objective of obtaining highest fidelity operation in the code space. 
For EsT gates, an appropriately weighted objective of error transparency (see Eq.~\eqref{eqn:app_et_cost_single},~\eqref{eqn:app_et_cost}) is added in addition to obtaining high fidelity operation in code and error subspaces.
To facilitate differentiability, we utilize a cost function based on state overlaps and fidelities rather than commutator norms and projectors for defining error transparency. 
Further details of the numerical optimization is described in App.~\ref{app:pulse_opt}.
Logical $X$ and $H$ gates are implemented with a duration of \SI{1}{\micro\second} with simultaneous qubit and cavity drives, while the $T$ gate is realized with only qubit drives in \SI{0.6}{\micro\second}.
To appropriately compare the performance of EsT and Ord gates, we select pulses that reach the same maximum active Fock state level, defined as the highest Fock state $|n\rangle$ to achieve a transient probability of occupation greater than $0.01$. 

\section{Results}

We simulate the dynamics of the oscillator under the Ord and EsT gates and calculate the violation metrics defined in Eq.~\eqref{eqn:Delta_qec}-\eqref{eqn:traj_mismatch}. 
We observe the optimization for EsT gate successfully minimizes violation of the QEC conditions in the instantaneous code space, reducing leakage out of the instantaneous error space due to photon loss, and enforcing trajectory and coherence matching in the instantaneous code and error spaces. 
The Ord gate has no such constraint and therefore exhibits substantial QEC violation, leakage and trajectory mismatch due to photon loss. 
We compare the QEC violation, leakage of the evolved error space out of the instantaneous error space during the operation and the trajectory mismatch in Fig.~\ref{fig:fig1}d-e respectively. 
Trajectory mismatch is plotted for evolution of initial state $|0_L\rangle$.
Values closer to zero suggest a lower violation of the instantaneous conditions required to satisfy ET.
These plots confirm how the optimizer enables the pulse to be EsT.

While these metrics quantify ET at a specific moment, the final gate fidelity will depend on both the immediate impact of the error and the subsequent evolution under the remaining control pulse.
To capture this full effect, we can evaluate the gate fidelity conditioned on a single photon loss event occurring at an arbitrary time during the operation.
As detailed in App.~\ref{app:numerical_simulation}, the EsT gate maintains a higher conditional fidelity across the pulse duration compared to Ord gate.

To verify the ET property, we performed Wigner tomography on the states following the application of 5 $X$ gates on the prepared state $\left|-Y_{L}\right\rangle  =(|0_L \rangle - i|1_L\rangle)/\sqrt{2}$. 
Fig.~\ref{fig:fig2}a shows ideal Wigner functions of initial (solid box) and final state (dashed box) after application of $X$ gate in the code and error space logical Bloch spheres.
We show the pulse sequence for the experiment in Fig.~\ref{fig:fig2}b.
After applying the gate, we read out the qubit state followed by parity measurement of the cavity state and finally perform Wigner tomography \cite{PhysRevLett.78.2547}.
If there is a photon loss error during the operation, the state ends in the error subspace.
Based on the parity measurement before the Wigner tomography, we post-select the final state in code and error subspaces.
We show the measured Wigner functions for prepared $\left|-Y_{L}\right\rangle$ states and the states in code (left) and error (right) space after EsT and ord gates in Fig.~\ref{fig:fig2}c. 
Code and error space Wigners have different fixed global frame rotation due to qubit being in the excited state for longer during the parity measurement process, which can be fixed with a frame rotation in software. 

The Wigner functions measured in the codespace indicate that the $X$ gate is accomplished and the logical states exhibit similar coherence within this subspace for both the EsT and Ord gates.
Crucially, phase coherence and state amplitudes in the error space are preserved for the EsT gate, whereas they are significantly corrupted for the Ord gate.

We quantify the performance in the code and error space conditioned on single photon loss by performing quantum process tomography (QPT) \cite{NielsenChuang2010, PRXQuantum.6.030202} of the states following the sequence shown in Fig.~\ref{fig:fig3}a and post-selecting on the code and error space states based on the parity measurement.
Different decode operations are used that respectively map code and error space states to ancilla qubit states for extracting the fidelities in those subspaces.
To focus only on the oscillator photon loss error to which the gate construction is EsT, we also post-select only the cases where the qubit is in the ground state at the end of the gate operations.
Details of this analysis and further experimental data are given in App.~\ref{app:add_expt}.

\begin{figure}[t]
\includegraphics[scale = 0.9]{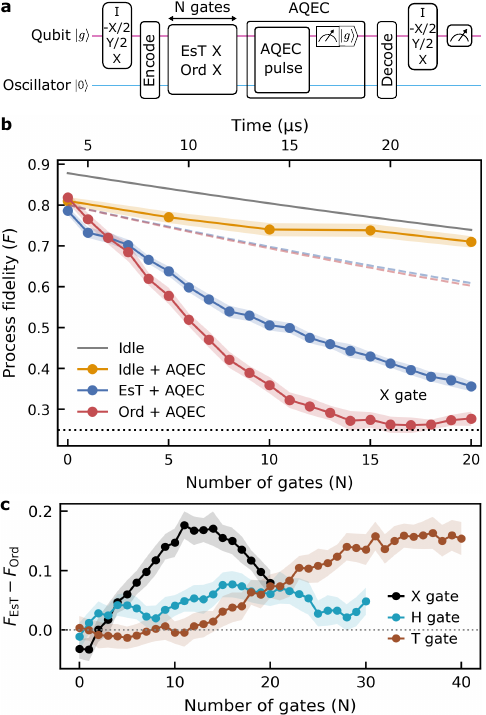}
\caption{\label{fig:fig4}\textbf{EsT gates protected by AQEC:}
(a) Experimental sequence to check gate performance with AQEC. 
We perform a single AQEC step after $N$ gates and perform process tomography. 
(b) We show the performance of EsT and Ord X gate with number of gates applied before doing error correction. 
EsT gate outperforms the Ord gate and the improvement increases with number of gates and then decreases as uncorrectable qubit errors, two photon loss errors in oscillator and control errors start to dominate. 
The plot also shows curves for process fidelity decay for idling and for applying an AQEC pulse after idling for an equivalent duration of N gates.
The blue and red dashed lines provide a visual reference for simulated idling decay curves, calculated at an average photon number equal to the average photon number during the gate operation.
The horizontal dashed line at process fidelity of 0.25 indicates a completely incoherent process and shaded regions around the data points signify 1 standard deviation of fidelities for repeated measurements.
The top X axis shows the total duration of the experiment that includes encode, decode, AQEC and active reset of qubit.
(c) Summary of improvement in gate fidelity for EsT gate compared to Ord gate by applying AQEC as a function of the number of gates for X, H and T gates. 
}
\end{figure}

Experimentally measured fidelities for the code (circles) and error (triangles) space are shown for EsT and Ord $X$ gates in Fig.~\ref{fig:fig3}b. 
To understand the dynamics and extract gate fidelities, we model the fidelity performance of the EsT and Ord gates as a function of the number of gates applied. 
The modeled curves, shown as dashed lines for code space and solid lines for error space, match our experimental data well.
This plot confirms that the fidelity in code space is comparable for EsT and Ord gates, whereas in the error space, EsT gates exhibit substantially higher fidelity compared to Ord gates.

The decay of error-space fidelity with gate depth appears non-intuitive at first glance. 
This is due to two processes that dominate over photon loss during the gates at short times (after only a few gates).
The first is due to state preparation, which has an error that produces some weight in error subspaces.
The second is that the parity measurement process itself has an error of $\sim$5\%.
So, at short times, a state selected in the error subspace has a small probability of being there as a result of photon loss during the gate.
We model these error channels to fit the observed decay curves, allowing us to estimate the expected decay profiles free of parity selection errors. (see App.~\ref{app:espace_model} for details).
These `intrinsic' fidelities of the gates are presented in Fig.~\ref{fig:fig3}c.
This plot further validates the similarity in code space performance and the improvement in error space performance for EsT gates compared to Ord gates.

\begin{figure*}[t]
\includegraphics[scale = 1]{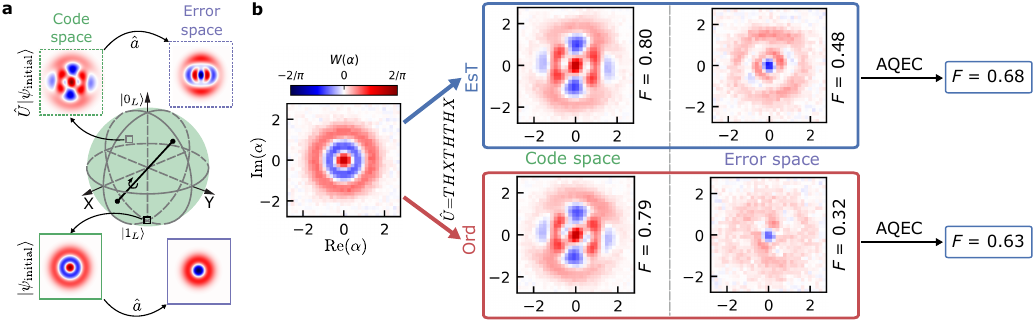}
\caption{\label{fig:fig5}\textbf{EsT protection for non-trivial sequence:}
(a) Logical code space Bloch sphere showing the ideal Wigner functions before (solid box) and after (dashed) the non-trivial $THXTHTHX$ operation sequence that effectively rotates the state around the marked non-trivial rotation axis (black line).
The ideal error space Wigner functions are also shown. 
(b) We show the experimentally measured Wigner functions of the prepared $|1_L\rangle$ state, the final state post-selected in code and error subspace after applying the pulse sequence comprising of EsT and Ord gates.
We measure the process fidelities as reported on the side of the measured Wigner functions.
Finally, we apply the AQEC pulse and measure the process fidelity. 
We observe a 0.05 improvement in process fidelity for the EsT gate sequence compared to the Ord gate sequence.
}
\end{figure*}

From the model that fits experimental data, we can extract the per-gate fidelity in case of a photon loss during the operation ($F_{\mathrm{err|jump}}$) and the fidelity of the operation in the error space ($F_{\mathrm{err}}$). 
For $X$ gate, $F_{\mathrm{err|jump}}$ increases from $0.45$ to $0.85$ and $F_{\mathrm{err}}$ increases from $0.4$ to $0.95$ for the EsT gate compared to Ord gate.
We summarize these numbers in Fig.~\ref{fig:fig3}d for $X$, $H$ and $T$ gates.
Averaged across the set of gates, we observe 5 times reduction in infidelity for the case of photon loss during the operation for the EsT gate compared to the Ord gate. 
This improvement is achieved because of the error transparency optimization.
The per-gate infidelity in error space is reduced 17 times for the EsT gate. 
This improvement is because the Ord gate optimization does not optimize for the fidelity of operation in error space, whereas the EsT gate is optimized for that as well.
This also raises the question whether optimizing for operation fidelity in both code and error subspaces is sufficient, without explicit error transparency optimization.
To investigate this, we introduce the LE gate, where `LE' stands for ``Logical and Error space" gate. 
The LE gate maximizes operation fidelity in the code and error subspaces separately, but does \textit{not} optimize for error transparency (see App.~\ref{app:pulse_opt} for details of pulse construction).
This gate shows intermediate performance that surpasses the Ord gate but remains inferior to the EsT gate. 
Additional experimental characterization and simulations are provided in App.~\ref{app:add_expt} and App.~\ref{app:numerical_simulation}, respectively. 

To demonstrate that the EsT gate performance improvement in error space remains valid after recovery to code space, we investigate the performance of gates under autonomous quantum error correction (AQEC) protection by applying the experimental sequence in Fig.~\ref{fig:fig4}a. 
The AQEC pulse is numerically optimized with a duration of \SI{1}{\micro\second} to recover error states to code states and transfer the error entropy to the ancilla qubit (see App. \ref{app:pulse_opt_AQEC}) . 
The ancilla qubit is then conditionally reset to the ground state, followed by a measurement which takes another \SI{1.2}{\micro\second} \cite{Ma2020}. 
We note that other AQEC methods have also been demonstrated in recent experiments, which can potentially perform better than this AQEC technique \cite{Gertler2021, 2510.19794, 2509.26042, 2509.22191}.

As shown in Fig.~\ref{fig:fig4}a, we repeatedly applied $N$ gates followed by a single AQEC step. 
Experimentally measured process fidelities for the EsT and Ord $X$ gates are shown in Fig.~\ref{fig:fig4}b.
As the gate depth $N$ increases, the difference in process fidelity between the EsT and Ord gates widens. 
The improvement comes from both being EsT and better performance of the EsT gate in the error subspace. 
As the number of gates increases, uncorrectable two-photon loss errors in the oscillator, ancilla qubit errors and control errors start to dominate, reducing the performance improvement.
Experimentally measured process fidelities for applying a single AQEC operation after waiting for a time equivalent to the gate operations are also plotted in Fig.~\ref{fig:fig4}b alongside a process fidelity decay curve for idling.
These curves demonstrate that while our AQEC pulse lowers the absolute measured fidelity, it actively recovers quantum information from the error space and thereby slows the decay of encoded, actively manipulated quantum information in code space. 
We have also added red and blue dashed lines as a visual reference for simulated idling decay curves, calculated at an average photon number equal to the average photon number during the gate operation.

We perform this experiment for $H$ and $T$ gates and summarize the improvement in fidelity for the EsT gates compared to the Ord gates in Fig.~\ref{fig:fig4}c.
For $X$ and $H$ gates, the fidelity improvement increases with the number of gates and then decreases due to the aforementioned reasons.
For T gate, we see performance improvement up to 40 gates and expect it to decrease after that.
We note here that EsT $H$ gate has a lower fidelity improvement with AQEC even though it had better error space performance fidelities ($F_{\mathrm{err|jump}}$ and $F_{\mathrm{err}}$) compared to the $X$ gate. 
We attribute this to the following two facts. 
Due to fluctuations of device parameters, EsT gate fidelity without the AQEC was lower for the $H$ gate, and the probability of oscillator photon loss during the gate is higher for the $H$ gate compared to the $X$ gate (see App.~\ref{app:add_expt} for details).
This could be resolved in the future by engineering robustness objectives into the numerical optimization~\cite{PhysRevApplied.17.014036, PhysRevLett.132.193801}. 
Control errors mentioned above may also be mitigated by calibration of pulses with hardware feedback using reinforcement learning methods \cite{PhysRevX.12.011059, PRXQuantum.2.040324}.

Finally, we demonstrate that the EsT property is preserved when gates are concatenated by performing a non-trivial unitary with the gate sequence ($THXTHTHX$) on a prepared $|1_L\rangle$ state.
This gate does an effective clockwise rotation of 158 degrees around the rotation axis (0.47, 0.75, 0.47) as shown on the code space Bloch sphere in Fig.~\ref{fig:fig5}. 
Ideal Wigner functions of the initial state (solid box) and final state (dashed box) in code and error space are also shown.
Fig.~\ref{fig:fig5}b shows the experimentally measured Wigner function of the prepared $|1_L\rangle$ state. 
After applying the gate sequence, following the protocol shown in Fig.~\ref{fig:fig2}b, we post-select our states in code and error spaces based on parity measurement and measure Wigner functions of states in those subspaces.
Both gates perform the desired operation with similar fidelity in code space. 
In the error space, the EsT gate preserves the phase coherence and state amplitudes of the target state better than the Ord gate.
Fidelities beside the Wigners are experimentally measured process fidelities in code and error subspaces.
Upon applying AQEC, we recovered process fidelities of $0.68$ and $0.63$ respectively for EsT and Ord gate, representing a net fidelity improvement of $0.05$ for the operation sequence.

We also simulate this sequence to understand the fidelities without SPAM and control errors,
We get code space fidelities of $0.94$ and $0.92$ for EsT and Ord gates sequence, respectively.
In the error space conditioned on single photon loss of the oscillator, the EsT sequence achieves a fidelity of 0.82 compared to 0.35 for Ord sequence. 
This shows a 3.6 times reduction in infidelity in the error space. 
Following AQEC, the final fidelities are 0.86 for EsT and and 0.82 for Ord sequence.
This confirms that the optimized gates are mutually compatible, enabling synthesis of universal control over oscillator encodings with discrete EsT operations. 

\section{Discussion and Outlook}
In this work, we design an error semi-transparent(EsT) operation with simple linear drives leveraging dynamic subspaces to have universal control on a binomial bosonic logical qubit. 
We show that trading off exact ET for simpler control can enable substantial performance improvement for actively manipulated bosonic quantum information.
Our design framework for EsT gates can be readily extended to other rotation-symmetric codes, such as four-legged cat codes \cite{Mirrahimi_2014, Ofek2016} and higher order binomial codes \cite{PhysRevX.6.031006}.
Our optimization framework also provides a pathway to explore the trade-offs for an approximate Eastin-Knill theorem \cite{PhysRevX.10.041018} in standard qubit architectures and construction of fast, approximately bias-preserving operations for erasure qubits \cite{PhysRevX.13.041022, PhysRevResearch.7.013249}. 

However, because the linear drive does not commute with the photon loss operation on the codespace, an exactly ET operation is not likely to be achievable.
In App.~\ref{app:lin_drive_limitation}, we describe further intuition for the limitations of the linear drive. 
Improvements in ET performance can also be obtained with optimization for additional objectives of minimizing mean photon number difference of evolving code words and leakage (see App.~\ref{app:lin_drive_limitation} for details).
An optimization involving dynamics obtained from solving the master equation, taking into account the loss rates of the system, will be able to account for the probability of photon jump and no-jump errors and therefore show further performance improvement. 
One could leverage dynamic subspace ET construction and obtain a theoretically exact ET with strong nonlinear drives (such as squeezing) using additional or different circuit elements. 
However, the utility of such schemes must be balanced with the increase in circuit and protocol complexity that potentially leads to more error channels. 

The remaining dominant error source in the circuit is the decoherence of the transmon ancilla.
Historically, the ancilla has had higher decoherence rates than the oscillator.
However, due to the use of higher oscillator levels during a fast amplitude-mixing gate, the effective error probability during the gate is comparable for both the oscillator and the qubit as measured in experiments (App.~\ref{app:add_expt}) and validated by simulations (App.~\ref{app:numerical_simulation}). 
In principle, qubit errors could be mitigated by replacing the qubit drives with a few simultaneous off-resonant Photon-Assisted Stark Shift (PASS) drives \cite{Ma2020}, optimizing their amplitude to impart the necessary geometric phase to the cavity states without exciting the ancilla. 
However, this approach necessitates inefficiently long gate times to maintain adiabaticity or risks higher decoherence rates and ancilla population with stronger drives.
Alternatively, a multi-level ancilla could be utilized to numerically enforce ancilla path independence alongside oscillator error transparency \cite{Reinhold2020, PhysRevLett.125.110503, PhysRevX.14.031016}. 
Utilizing the $g$-$f$ transition of the ancilla for control would facilitate the detection of ancilla errors by monitoring the leakage to the state $|e\rangle$, thereby heralding ancilla errors \cite{PRXQuantum.4.020354, 2601.21838}.
Such heralding of error is advantageous as it converts generic errors into erasure errors with known location. 

In summary, we have introduced the concept of dynamic subspaces as a framework for realizing Error semi-transparent (EsT) operations. 
By traversing beyond static code and error spaces, this approach enables fast universal control of bosonic encodings using simpler linear drives. 
Our results demonstrate that EsT operations significantly enhance the fidelity of actively controlled bosonic qubits.
The EsT gate set can be readily used for enhancing circuit depth for sensing and algorithmic applications and quantum signal processing~\cite{PRXQuantum.2.040203}.
Our approach is also compatible with methods for ancilla error detection  \cite{PRXQuantum.4.020354, 2601.21838} offering a practical pathway towards error-mitigated universal control of bosonic quantum information. 

\begin{acknowledgments}
We acknowledge B. Cole, C. P. Larson, E. Yelton, B. L. T. Plourde, and Luojia Zhang for help in fabrication of the chip with the transmon and readout resonator and Christopher S. Wang for design of the superconducting cavity.
We thank Haoran Lu and Kushagra Aggarwal for their help with the cryogenic microwave setup. 
We also thank Maciej Olszewski and Xiangqin Wang for helpful discussions.
We acknowledge helpful discussions and comments on the manuscript by Baptiste Royer. 
We thank Sridhar Prabhu, Peter McMahon, Chen Wang, and Christopher S. Wang for valuable comments on the manuscript.
V.F. thanks I. L. Chuang for recommending a quantum signal processing lens.

This work was supported by the NSF under award number PHY 2512537, as well as by the Aref and Manon Lahham Faculty Fellowship. 
We gratefully acknowledge Nord Quantique for the fabrication of the $\lambda/4$ superconducting cavity and MIT Lincoln Laboratory for supplying the Josephson traveling-wave parametric amplifier used in our experiments.
This work was performed in part at the Cornell NanoScale Facility, a member of the National Nanotechnology Coordinated Infrastructure (NNCI), which is supported by the National Science Foundation (NSF) (Grant No. NNCI-2025233).

\textbf{AI usage disclosure} We have used Google Gemini 3 model strictly to assist with stylization and formatting of the figures only. 

\textbf{Author contributions} S. R. and V. F. proposed the idea of using linear drives for EsT operation. 
S. R. developed the dynamic subspace EsT theory with O. C. W. and V. F..
S. R. performed the numerical optimizations, experiments, and analysis. 
S. R. and V. F. wrote the manuscript with input from O. C. W.. 
V. F. supervised the project.

\end{acknowledgments}

\section*{Data and code availability}
Data and code that support the findings of this article are openly available in 10.5281/zenodo.18916619.

\newpage
\appendix
\clearpage
\begin{center}
\textbf{\large Appendices}
\end{center}
\setcounter{figure}{0} 
\renewcommand{\thefigure}{A\arabic{figure}}
\renewcommand{\theHfigure}{A\arabic{figure}}

\section{Theory}
\label{app:lin_drive_limitation}

\textbf{Limitations of linear drive:}
Here we analyze the limitations of purely linear drives to achieve gates ET to photon loss in the oscillator. 
In the dispersive regime, the coupled cavity-qubit Hamiltonian when driven can be written as 
\begin{align}
\label{eqn:app_full_ham_begin_1}
H(t) &= H_c + H_q + H_{\text{int}} + H_{\text{drive}}(t)~, \\
H_a/\hbar &= \omega_a\hat{a}^\dagger\hat{a} + \frac{K}{2}(\hat{a}^\dagger)^2\hat{a}^2 + \frac{K'}{6}(\hat{a}^\dagger)^3\hat{a}^3~, \\
H_q/\hbar &= \omega_q\hat{q}^\dagger\hat{q} + \frac{K_q}{2}(\hat{q}^\dagger)^2\hat{q}^2~, \\
H_{\text{int}}/\hbar &= \chi\hat{a}^\dagger\hat{a}\hat{q}^\dagger\hat{q} + \frac{\chi'}{2}\hat{q}^\dagger\hat{q}(\hat{a}^\dagger)^2\hat{a}^2~, \\
H_{\text{drive}}(t)/\hbar &= \epsilon(t)\hat{a} + \Omega(t)\hat{q} + \text{h.c.}
\label{eqn:app_full_ham_1}
\end{align}
where, $\omega_a$ and $\omega_q$ are the angular frequencies and $\hat{a}$ and $\hat{q}$ are the corresponding lowering operators of the oscillator and the qubit, respectively; $K$ is the self-Kerr nonlinearity of the oscillator, $K^{\prime}$ is the higher order self-Kerr of the oscillator, and $K_q$ represents the anharmonicity of the qubit. 
The system interaction is characterized by the dispersive coupling strength $\chi$ and the higher-order coupling term $\chi'$. 
Finally, $\epsilon(t)$ and $\Omega(t)$ denote the time-dependent complex drive amplitudes applied to the oscillator and qubit ports.
Even though the higher order terms were not described in the main text, it is important to take them into account for an accurate description of the system. 

Considering the full Hamiltonian in Eq.~\eqref{eqn:app_full_ham_begin_1}-\eqref{eqn:app_full_ham_1}, we see that the commutator of terms like $\omega_a\hat{a}^{\dagger}\hat{a}$, $\omega_q\hat{q}^{\dagger}\hat{q}$ and $\chi\hat{a}^{\dagger}\hat{a}\hat{q}^{\dagger}\hat{q}$ with the oscillator photon jump operator ($\hat{a}$) are proportional to $\hat{a}$.
Consequently, a photon loss event during evolution under these terms of the Hamiltonian maps the final logical state $|\psi_L\rangle$ to the corresponding error space state $\hat{a}|\psi_L\rangle$, up to a deterministic phase.
More generally, any terms of the Hamiltonian whose commutators with $\hat{a}$ are correctable will map a photon loss event to a correctable error under evolution by those terms (this is known as error-closure~\cite{PRXQuantum.4.020354}, a generalization of error-transparency).

However, we identify two terms whose commutators with the photon loss operator ($\hat{a}$) are not correctable: the oscillator self-Kerr non-linearity ($K\hat{a}^{\dagger}\hat{a}^{\dagger}\hat{a}\hat{a}/2$) and the higher order dispersive shift ($\chi^{\prime}\hat{a}^{\dagger}\hat{a}^{\dagger}\hat{a}\hat{a}\hat{q}^\dagger\hat{q}/2$):
\begin{equation}
    \label{eqn:app_ham_commute}
    \bigg[ \frac{K}{2} \hat{a}^{\dagger}\hat{a}^{\dagger}\hat{a}\hat{a} + \frac{\chi^{\prime}}{2}\hat{a}^{\dagger}\hat{a}^{\dagger}\hat{a}\hat{a}\hat{q}^\dagger\hat{q}, \hat{a} \bigg]= -(K + \chi^{\prime}\hat{q}^{\dagger}\hat{q})\hat{a}^{\dagger}\hat{a}^{2} ~.
\end{equation}
The nonlinear terms $K$ and $\chi^{\prime}$ introduce an amplitude-dependent differential rotation between the code and error subspaces. 
This can somewhat be mitigated by an appropriate linear drive $\epsilon(t)$ at each time point, but can not be fully mitigated as the linear drive lacks the non-linear operations required to fully correct the shearing error. 

Eq.~\eqref{eqn:app_ham_commute} dictates a fundamental trade-off. 
To improve ET fidelity, we can engineer hardware with minimal $K$ and $\chi^{\prime}$ by decreasing the dispersive coupling $\chi$ of the cavity and the qubit.  
One can also minimize $\epsilon(t)$ and optimize the trajectory to lower intermediate photon number ($\bar{n}$) during the evolution. 
However, such weaker coupling and drive strengths necessitate longer gate duration ($t_g)$, which increases the integrated probability of photon loss ($\kappa_c\bar{n}t_g$) and exposes the system to errors from other sources (e.g., qubit decoherence), where $\kappa_c = 1/T_{1,c}$ is the oscillator  photon loss decay rate.

\textbf{Possible improvements:}
Further the linear drive is insufficient to guarantee fidelity under deterministic no-jump evolution. 
In the Lindblad master equation, the continuous non-unitary back-action is governed by the effective non-Hermitian Hamiltonian $\hat{H}_{\text{eff}} = \hat{H} - i\frac{\kappa_c}{2}\hat{a}^{\dagger}\hat{a}$.
Because our code words span superposition of different Fock states, the state-dependent amplitude damping deterministically distorts the code space, even in the absence if discrete jump errors. 
This continuous distortion can be captured and optimized by simulating the full master equation or stochastic Schrodinger equation to improve performance of the gate operation. 
However, such pulse optimization is computationally more intensive than using the Schrodinger equation and remains vulnerable to experimental fluctuations in $\kappa_c$.

A secondary limitation of our current protocol is that it does not directly minimize $\Delta_{\text{QEC}}$. 
In the absence of this logical states will undergo amplitude damping at different rates.
More importantly, during the photon loss, the environment can gain information about `from which state the photon loss is occurring' and thereby dephase the logical information. 
This is an irreversible error that the standard ET protocol can not correct at the end of the operation. 
However, we can take this into account with another optimization objective in the optimal control problem and improve the gate fidelity. 
We can minimize $\Delta_{\text{QEC}}$ directly or specifically for the binomial `kitten' code by minimizing $\Delta\bar{n}_L(t)$, defined as
\begin{equation}
\begin{split}
    \Delta \bar{n}_L(t) =& \langle 0_L | \hat{U}^{\dagger}(t,0) \hat{a}^\dagger \hat{a} \hat{U}(t,0) |0_L \rangle \\
    &- \langle 1_L | \hat{U}^{\dagger}(t,0) \hat{a}^\dagger \hat{a} \hat{U}(t,0) |1_L \rangle~.
\end{split}
\end{equation}
Similarly, minimizing the leakage out of instantaneous error subspace can also be added as an additional objective to improve error transparency of the operation. 

\textbf{QEC violation metric derivation:}
To quantify violation of the Knill-Laflamme condition in our dynamic code space, we adapt the quantum error-correction (QEC) matrix formalism used by Albert \textit{et al.} in Ref.~\cite{PhysRevA.97.032346}. 
For a general error channel characterized by Kraus operators $\{E_k\}$, the Knill-Laflamme QEC conditions necessitate that the projection of any error product $E_i^\dagger E_j$ onto the code space is proportional to the logical identity. 
For a dynamic code space with instantaneous projector $P_C(t) = |0_C(t)\rangle\langle 0_C(t)| + |1_C(t)\rangle\langle 1_C(t)|$, the effect of these errors is captured by the $2 \times 2$ instantaneous QEC matrices:
\begin{equation}
    M_{ij}(t) = P_C(t) E_i^\dagger E_j P_C(t)~.
    \label{eqn:app_qec_matrix}
\end{equation}
Because $P_C(t)$ serves as the identity operator within the two-dimensional logical subspace, $M_{ij}(t)$ can be uniquely expanded in the instantaneous logical Pauli basis $\{P_C(t), X_C(t), Y_C(t), Z_C(t)\}$:
\begin{equation}
\begin{split}
    M_{ij}(t) &= c_{ij}(t) P_C(t) + x_{ij}(t) X_C(t) \\
    &\quad + y_{ij}(t) Y_C(t) + z_{ij}(t) Z_C(t)~,
\end{split}
\label{eqn:app_pauli_decomp}
\end{equation}
where the complex coefficients are extracted via the trace inner product: $[c,x,y,z]_{ij}(t) = \frac{1}{2}\text{Tr}\left\{ [P,X,Y,Z]_C(t) M_{ij}(t) \right\}$. 

Exact error correction requires $M_{ij}(t) = c_{ij}(t)P_C(t)$ for all error combinations. 
The non-identity coefficients $x_{ij}(t)$, $y_{ij}(t)$, and $z_{ij}(t)$ quantify the magnitude of uncorrectable logical bit, phase, and joint bit-phase flips. 

To construct a scalar metric $\Delta_{\text{QEC}}(t)$ that captures the total deviation from exact correctability, we isolate these uncorrectable parts. 
Because the logical Pauli matrices $X_C, Y_C, Z_C$ are strictly traceless and $\text{Tr}[P_C(t)] = 2$, taking the trace of Eq.~\eqref{eqn:app_pauli_decomp} yields $\text{Tr}[M_{ij}(t)] = 2c_{ij}(t)$. 
Subtracting $\frac{1}{2}\text{Tr}[M_{ij}(t)]P_C(t)$ from $M_{ij}(t)$ isolates the uncorrectable error generators:
\begin{equation}
\begin{split}
    M_{ij}(t) - \frac{1}{2}\text{Tr}\left[M_{ij}(t)\right] P_C(t) &= x_{ij}(t) X_C(t) \\
    &\quad + y_{ij}(t) Y_C(t) + z_{ij}(t) Z_C(t)~.
\end{split}
\end{equation}
We quantify the magnitude of this uncorrectable component using the Frobenius norm, $\|A\|_2 = \text{Tr}(A^\dagger A)$. 
Utilizing the algebraic properties of the Pauli matrices, all cross-terms vanish under the trace. The norm evaluates directly to the sum of the squared error magnitudes:
\begin{equation}
\begin{split}
    \Delta_{\text{QEC}}(t) = &\left\| M_{ij}(t) - \frac{1}{2}\text{Tr}\left[M_{ij}(t)\right] P_C(t) \right\|_2 \\
    &\quad = 2 \left( |x_{ij}(t)|^2 + |y_{ij}(t)|^2 + |z_{ij}(t)|^2 \right).
\end{split}
\label{eqn:app_frobenius_eval}
\end{equation}
Summing this squared distance over all relevant error combinations $i, j$ yields the total QEC violation metric $\Delta_{\text{QEC}}(t)$ presented in the main text. 

For the bosonic codes, the relevant error generators are $\{ \hat{a}, \hat{n}\}$ which correspond to single photon loss in the oscillator and dephasing, backaction respectively. 
Thus the above condition  can be summed for $\mathcal{O} \in \{\hat{a}, \hat{n}\}$. 
Expressing the Pauli coefficients in terms of matrix elements $\mathcal{O}_{\mu\nu}(t) = \langle \mu_C(t)| \hat{\mathcal{O}} | \nu_C(t) \rangle$, the metric simplifies to the scalar summation
\begin{align}
    \Delta_{\text{QEC}}(t) &= \sum_{\mathcal{O} \in \{\hat{a}, \hat{n}\}} \left\| M_{\mathcal{O}}(t) - \frac{1}{2}\text{Tr}\left[M_{\mathcal{O}}(t)\right] P_C(t) \right\|_2 .
    \label{eqn:app_scalar_qec_metric}
\end{align}
This formulation provides a computationally and analytically accessible cost function for bounding the QEC violation.

\section{Experimental setup}
\label{app:expt_setup}

\begin{table*}[t]
    \begin{center}
    \begin{tabular}{ |l|c|c|c| } 
    \hline
    \textbf{Quantity} & \textbf{Symbol} & \textbf{Measured value} & \textbf{Hamiltonian term}\\
    \hline \hline
    Cavity frequency & $\omega_a / 2\pi$ & \SI{6.05}{\giga\hertz} & $\omega_c a^ \dagger a$\\
    \hline
    Readout frequency & $\omega_r / 2\pi$ & \SI{8.92}{\giga\hertz} & $\omega_r r^ \dagger r$\\
    \hline
    Qubit frequency & $\omega_q / 2\pi$ & \SI{5.60}{\giga\hertz} & $\omega_q q^ \dagger q$\\
    \hline
    Qubit-readout dispersive coupling & $\chi_r / 2\pi$ & \SI{-600}{\kilo\hertz} & $\chi_r r^\dagger r |e\rangle \langle e|$ \\
    \hline
    Qubit-Cavity dispersive coupling & $\chi/2\pi$ & \SI{-3.66}{\mega\hertz} & $\chi a^\dagger a |e\rangle \langle e|$\\
    \hline
    Qubit anharmonicity & $K_q/2\pi$ & \SI{-180}{\mega\hertz} & $K_q q^\dagger q^\dagger q q /2$\\
    \hline
    Second order cavity-qubit dispersive coupling & $\chi^\prime/2\pi$ & \SI{39}{\kilo\hertz} & $\chi^\prime a^\dagger a^\dagger a a |e \rangle \langle e|/2$\\
    \hline
    Cavity Self-Kerr & $K/2\pi$ & \SI{-22}{\kilo\hertz} & $K a^\dagger a^\dagger a a /2$\\
    \hline
    Second order Cavity Self-Kerr & $K'/2\pi$ & \SI{590}{\hertz} & $K' a^{\dagger 3} a^3 /6$\\
    \hline
    Qubit lifetime & $T_{1q}$ & \SI{70}{\micro\second} & \\
    \hline
    Qubit dephasing time, Ramsey &  $T_{2q,R}$ & \SI{30}{\micro\second} & \\
    \hline
    Qubit dephasing time, Hahn-echo &  $T_{2q,E}$ & \SI{84}{\micro\second} & \\
    \hline
    Cavity lifetime & $T_{1a}$ & \SI{180}{\micro\second} & \\
    \hline
    Cavity dephasing time & $T_{2a,R}$ & \SI{290}{\micro\second} & \\
    \hline
    Qubit thermal population & & 1.5\% & \\
    \hline
    Cavity thermal population & & 0.9\% & \\
    \hline
    \end{tabular}
    \end{center}
    \caption{\textbf{Device parameters:} Device parameters measured with standard spectroscopic and time-domain methods. 
    The qubit has some TLS associated with it which makes it respond at two different frequencies about 50-\SI{60}{\kilo\hertz} apart as seen in qubit spectroscopy and Ramsey experiments.
    }
    \label{tab:params_table}
\end{table*}

\begin{figure}[h!]
\centering
\includegraphics[scale = 1]{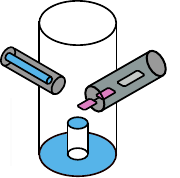}
\caption{\textbf{Schematic of the 3D package}: We use 3D $\lambda/4$ cavity made (light blue) of 4N Aluminum as the oscillator mode. 
An ancilla transmon qubit (pink) is coupled to the oscillator and used for controlling the oscillator. 
The transmon has a readout resonator (gray rectangle) coupled to it for readout of the transmon state.
}
\label{fig:app_device_schematic}
\end{figure}

Schematic design of the package is shown in Fig.~\ref{fig:app_device_schematic}. 
In this section we describe the production and characterization of the package, with measured quantities shown in Table.~\ref{tab:params_table}. 
This is the same package as used in Roy, \textit{et al}\cite{PhysRevX.15.021009} with minor changes in the cryogenic wiring.

We used a $\lambda/4$ cavity made of 4N Aluminum as the cavity mode~\cite{PhysRevB.94.014506} (harmonic oscillator) in which we encode our binomial logical qubit.  
The cavity was treated with an acid etch to remove surface impurities. 
While this process typically produces cavities with lifetimes approaching the millisecond scale, the surface appears to have degraded due to shipment, long shelf time exposed to air, resulting in the measured cavity lifetime of \SI{180}{\micro\second}. 

The superconducting qubit used is a transmon with Nb capacitor pads, and an Al/AlO\textsubscript{x}/Al Josephson junction fabricated on a high-resistivity ($>\SI{10}{\kilo\ohm\centi\meter}$) silicon chip. 
Following a strip of the native silicon oxide in 2\% hydrofluoric acid, a \SI{75}{\nano\meter} thick Nb film was sputtered at the rate of \SI{50}{\nano\meter\per\minute}.
This Nb was patterned using photo-lithography to make the capacitor pads of transmon and the readout resonator.    
The Josephson junction for the qubit was fabricated in a Dolan bridge process~\cite{dolan_offset_1977} with bilayer MMA/PMMA resist. 
Double-angle evaporation for the Al-AlO\textsubscript{x}-Al was accomplished to form the junction. Before the first Al deposition an in-situ ion mill was performed to clean the top surface of Nb. After the first Al deposition, AlO\textsubscript{x} layer was formed by oxidizing the Al surface for 30 minutes at 7.25 Torr oxygen pressure. After pumping out the oxygen from the chamber, the second layer of Al was deposited . 
The area of the deposited Josephson junction is \SI{0.047}{\micro\meter^2}.
The chip  with the transmon qubit and readout resonator is mounted in the 3D cavity with a copper clamp .

\begin{figure*}
\centering
\includegraphics[scale = 1.25]{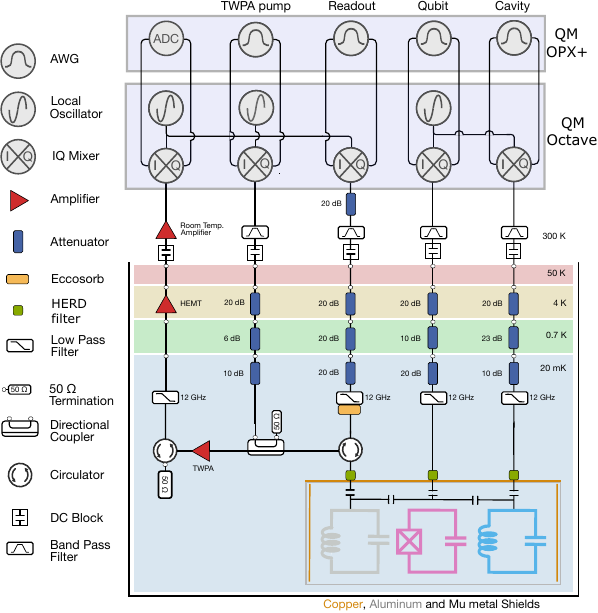}
\caption{\textbf{Experimental wiring diagram}: 
Schematic of room temperature and cryogenic microwave circuit.}
\label{fig:app_wiring}
\end{figure*}

The whole package is mounted on an oxygen free high conductivity (OFHC) copper bracket and has multiple layers of shielding, in order from inner to outer: (1) a Berkeley Black \cite{10.1063/1.1149739} coated copper shim to absorb any stray mm-wave radiation, (2) an aluminum can with an indium seal to an upper flange, which itself has indium-sealed SMA feedthroughs, (3) a mixing chamber can. 
This is depicted schematically in Fig.~\ref{fig:app_wiring}.
The dilution fridge also hosts a room-temperature magnetic shield that lines the vacuum can. 

The control pulses are generated using Quantum Machines OPX plus instrument and up-converted using local oscillator (LO) and IQ mixers in the Quantum Machines Octave. 
All pulses are digitally triggered with \SI{100}{\nano\second} buffer on the trigger on either side of the analog pulse.
The qubit and storage drives share the same LO whereas the readout up and down conversion share the same LO.  
Output of the readout signal is first amplified with a Travelling Wave Parametric Amplifier (TWPA) at base temperature, followed by a High electron mobility transistor (HEMT) at \SI{4}{\kelvin} and a room temperature amplifier (ZVA-1W-103+ Mini-Circuits). 
The TWPA is also pulsed with \SI{100}{\nano\second} buffer on both ends of the readout pulse.
Each of the input lines have K \& L low pass filter and Eccosorb filters. 
All the last attenuators, microwave filters, Eccosorb and HERD filters are connected to the 
copper bracket using a copper braid to improve thermalization.

\section{Pulse optimization method \label{app:pulse_opt}}

For achieving any operation on the joint cavity-qubit system, we optimize the complex time-dependent drive amplitudes $\epsilon(t)$ and $\Omega(t)$ that is simultaneously applied to the cavity and the qubit on resonance.
We simulate the system dynamics governed by the driven Hamiltonian $H(t)$ by numerically integrating the Schrodinger equation using the Dynamiqs package \cite{guilmin2025dynamiqs}. 
The pulse shapes are then optimized via the Gradient Ascent Pulse Engineering (GRAPE) method \cite{KHANEJA2005296}, utilizing JAX \cite{jax2018github} for automatic differentiation and the Adam optimizer \cite{DBLP:journals/corr/KingmaB14} from the Optax library \cite{deepmind2020jax} for gradient-based parameter updates.

We define a few different cost functions and use a single net weighted cost that the optimizer tries to minimize for parameter update. 
For easy computation of some cost functions, we simulate the evolution of states instead of simulating the propagator under the Hamiltonian. 
We start from six cardinal points in the Bloch sphere, evolve them and calculate fidelities and other costs based on the evolved state. 
We are interested in achieving high fidelity for all our operations. 
So starting from a set of initial states $|\psi_L\rangle$, we evolve the states under the time dependent Hamiltonian $H(t)$ to get the final states $|\psi_{\mathrm{final}}\rangle$ and then calculate the overlap fidelity with the target set of states. 
The cost function is defined as the infidelity
\begin{equation}
    C_1 = 1 - F_{\mathrm{target}} = 1 - \frac{1}{6}\sum_{i=1}^6 |\langle \psi_{\mathrm{final}, i} | \psi_{\mathrm{target}, i} \rangle|^2~,
    \label{eqn:app_fid_cost}
\end{equation}
where the sum is over the number of states used in the simulation.
We use the six cardinal points of the logical Bloch sphere in our simulation. 

Next, we consider enforcing dynamic error transparency. 
For this, we construct a second cost function based on the intermediate trajectory of the logical states based on the theory in Appendix~\ref{app:lin_drive_limitation}.
We evolve our states starting from cardinal points in the code and the error space, denoted by $\psi_C(t_0)$ and $\psi_{E_j}(t_0)$ respectively, where $t_0 = 0$. 
Constrained by the temporal resolution of the control hardware and computational overhead for optimization, we discretize the evolution generate by $H(t)$ into $N$ steps of resolution $\Delta t = t_i - t_{i-1}$ =  \SI{1}{\nano\second}.
We evaluate the instantaneous states $\psi_C(t_i)$ and $\psi_{E_j}(t_i)$ at the $i$-th time step by solving the Schrodinger equation with $H(t)$. 
We define a per step normalized ET fidelity metric using these evolving states as 
\begin{equation}
    \label{eqn:app_et_cost_single}
    F_{\text{ET}} = \frac{1}{N} \sum_{i=0}^{N-1} \frac{|\langle \psi_{E_j}(t_i) | \hat{a} | \psi_C(t_i) \rangle|^2}{\mathcal{N}}~,
\end{equation}
where, $\hat{a}$ is the bosonic annihilation operator and $\mathcal{N}$ is the normalization factor, $\mathcal{N} = \sqrt{\langle \psi_C(t_i)| a^\dagger a |\psi_C(t_i) \rangle}$.
Note that calculating ET fidelity this way is equivalent to partially satisfying all three ET condition described in Eq.~\eqref{eqn:exact_qec}-\eqref{eqn:ET_cond}.
This per step ET fidelity is then averaged over six set of starting points $(\psi_C(t_0), \psi_E(t_0))$ corresponding to cardinal points of the logical Bloch sphere. 
We call this $F_{\text{ET, avg}}$.
Then the corresponding cost function is written as 
\begin{equation}
    C_2 = 1 - F_{\text{ET, avg}}~.
    \label{eqn:app_et_cost}
\end{equation}

Eq.~\eqref{eqn:app_fid_cost} and \eqref{eqn:app_et_cost} are the two main cost functions used in this work.
For some initializations of $\epsilon(t)$ and $\Omega(t)$, we had to use a regularizer cost function 
% that we call `velocity variance' 
to ensure that the optimizer does not simply minimize the pulse dynamics duration to artificially inflate $F_{ET}$. 
For this, we can calculate the Fubini-Study distance (FSD) metric between consecutive steps, which is defined as 
\begin{equation}
    \text{FSD} = 2 \arccos{|\langle \psi_1| \psi_2\rangle|}~, 
\end{equation}
with $|\psi_1\rangle$ and $|\psi_2\rangle$ being the two states we want to calculate the distance between.
We can then define a velocity metric ($v_i$) using the Fubini-Study distance between two states at two consecutive states during evolution of our driven system, i.e., between $|\psi_C(t_i)\rangle$ and $|\psi_C(t_{i+1})\rangle$,
\begin{equation}
    v_i = (2/\Delta t) \arccos{|\langle \psi_C(t_i)| \psi_C(t_{i+1})\rangle|}~,
\end{equation}
We can then calculate the variance of this metric, 
\begin{equation}
    \sigma^2 = \frac{1}{N}\sum_{i=0}^{N-1} \left( v_i - \bigg( \frac{1}{N}\sum_{i=0}^{N-1} v_i \bigg) \right)^2~,
\end{equation}
Informally, we refer to this as `velocity variance' as this is the variance of a metric that calculates a distance per unit time. 
This metric is then averaged over six cardinal points as the initial state.
We define the cost function as 
\begin{equation}
    C_3 = \frac{1}{6}\sum_{j=1}^6 \sigma_j^2~.
\end{equation} 
This cost function ensures there is some dynamics at each step and thereby incentivizes the optimizer to not minimize the effective pulse duration.
This regularizer is a convenient way to ensure that operations of different types (e.g., EsT and Ord) have similar duration so that the opportunity for incoherent errors to influence the operation is comparable.
We also note that this cost function does not improve ET performance directly, and we acknowledge that other cost function metrics could accomplish a similar objective.

Finally we combine the cost functions with appropriate weights $w_i$ (manually tuned) to feed a net cost function for the optimizer to minimize. 
This is done as 
\begin{equation}
    C_{\text{tot}} = \sum_i w_i C_i~.
    \label{eqn:app_tot_cost}
\end{equation}
We do the optimization at two stages. 
A first stage optimization is performed with weights typically as $\vec{w}=(w_1, w_2, w_3) = (1, 0.6- 0.75, 5-10)$ for fidelity, ET and velocity variance.
This allows us to reach ET fidelity of up to $0.83$ and fidelity in the $0.9 - 0.99$ range.
Beyond this a second stage optimization with weights $\vec{w} = (1, 0.1, 0)$ are used to maximize the fidelity of the operation. 
At this step we notice ET fidelity and velocity variance change at around 1\% and less than 1\%, respectively.

We also add additional restrictions on the pulse shape. 
Due to the finite bandwidth of our arbitrary wave generator, we Fourier transform the pulse and remove any frequency components beyond the bandwidth of \SI{50}{\mega\hertz}. 
This helps to keep the pulses reliable and provide good matching of the simulation and experimental results. 
We also employ a maximum drive strength of $\Omega_{\text{max}}/2\pi = \epsilon_{\text{max}}/2\pi = $ \SI{4}{\mega\hertz}. 
Further, we enforce a \SI{48}{\nano\second} gaussian rise and fall duration of the pulse to ensure smooth ramp on and ramp off the pulse from and to zero drive amplitude.

By discretizing the control fields into \SI{1}{\nano\second} piecewise-constant intervals and applying these optimization methods, we obtain EsT gates with cost function in Eq.~\eqref{eqn:app_tot_cost} and Ord gates with cost function defined in Eq.~\eqref{eqn:app_fid_cost}. 
$X$ and $H$ gates are accomplished in \SI{1}{\micro\second} by simultaneously driving the qubit and cavity on resonance whereas, $T$ gate is achieved in \SI{0.6}{\micro\second} by driving the qubit only. 
Eq.~\eqref{eqn:app_et_cost} partially optimizes all three dynamic ET conditions described in Eq.~\eqref{eqn:exact_qec}-\eqref{eqn:ET_cond}.
This cost function was used for its computational simplicity in gradient calculation and the minimal weight tuning required within multi-objective optimization framework used.

We can also use additional cost functions: (a) one that minimizes the difference of average photon number $(\bar{n})$ between the evolving codewords to satisfy the dynamic QEC condition; one that enforces the equality of instantaneous and evolved error spaces; (b) one that keeps the pulse shapes smooth; (c) one that improve the robustness of the pulse to parameter and drive amplitude fluctuations. 
These can be employed in the same framework, but  careful tuning of the weights is needed to achieve desired goals. 
The weights can also be included as something to optimize but that requires a more sophisticated optimization method.
Techniques like projecting conflicting gradients, and conflict aware gradient optimization techniques may also be used for better optimization results.
Further using multi-level ancilla, ancilla ET can be similarly defined and included in the optimization directly. 
These will be explored in future works.

\subsection{LE gate}
\label{app:LE_gate}
In the main text we have described a standard non-ET gate (named Ord) gate, that is optimized for only the fidelity of the operation in the code space. 
The EsT gate then optimizes for fidelity of operation in both code and error space, and also for error transparency. 
In between these two optimization schemes, we can optimize for another operation where the fidelity of the operation in code and error space is maximized but the ET condition is not enforced. 
We name this LE gate. 
It is expected that this gate will have a better error space gate fidelity $F_{\text{err}}$ (as it is optimized for that) but will have worse fidelity error transparency for the case of photon loss during operation. 
So the performance of this gate will be somewhere between the Ord and EsT gate. 
This is verified by simulation in App.~\ref{app:numerical_simulation} and experimental results shown in App.~\ref{app:add_expt} for X gate.

\subsection{AQEC pulse}
\label{app:pulse_opt_AQEC}

Autonomous quantum error correction (AQEC) is equivalent to standard QEC consisting of error detection and correction operations.
AQEC is designed to remove the parity measurement and related errors and rely on simpler qubit projection. 
The pulse is designed to be an unitary operation that maps error space state $|\psi_E\rangle $ to codespace and transfers the entropy to ancilla excitation and keeps the code space state $|\psi_L\rangle$ undisturbed~\cite{Ma2020}.
\begin{align}
    U | \psi_{\mathrm{E}} \rangle | g \rangle &= | \psi_{\mathrm{L}} \rangle | e \rangle~, \\
    U | \psi_{\mathrm{L}} \rangle | g \rangle &= | \psi_{\mathrm{L}} \rangle | g \rangle~.
\end{align}
After applying this pulse, the qubit state is measured and a conditional unselective pi pulse is applied to bring the qubit in ground state unselective of the oscillator state.
This readout creates a phase difference between the code and error space which is taken into account in the pulse. 
We also take into account the evolution to self-Kerr, higher order self-Kerr, higher order dispersive shifts for the duration of the readout and ancilla reset to have a correct state at the end of the operation. 
We have numerically optimized the pulse shapes of drives simultaneously applied on the cavity and qubit to accomplish AQEC pulse in \SI{1}{\micro\second}. 
The pulse duration was chosen to be the shortest to minimize coherence related errors and control errors (that increase for shorter drive durations because of high amplitude, broad bandwidth pulses and complex dynamics of the system).

We can improve the performance of AQEC pulse by improving the following aspects: 
(1) Shorter readout with lower photon numbers by improving the coupling of the readout and the qubit,
(2) Readout pulse shaping, such as CLEAR pulse~\cite{PhysRevApplied.5.011001}, that decays all photon out of the resonator shortening the effective duration of active reset,
(2) Adjusted $\chi$ of qubit and cavity that balances gate speed and finite bandwidth issues of the short pi pulse,
(3) Improving on control errors with reinforcement learning based hardware-feedback controls~\cite{PhysRevX.12.011059, PRXQuantum.2.040324}, 
(4) improving coherence times of the cavity and qubit and 
(5) Employing other autonomous error correction methods \cite{Gertler2021, 2510.19794, 2509.26042, 2509.22191}.

\subsection{Encode-Decode operations}

We designed encode operations that take any state in the qubit Bloch sphere to logical code space Bloch sphere. 
This pulse is used to create logical code space states. 
We also design decode operations that take logical Bloch sphere states to qubit Bloch sphere. 
Then the qubit state can be probed to generate the density matrix of the state and full process tomography can be performed in this way. 
For probing code and error space processes, we generate two different decode pulses for code and error space. 
These pulses are \SI{0.6}{\micro\second} long.

\section{Simulation results}
\label{app:numerical_simulation}
\begin{figure*}[t]
\includegraphics[width = 0.9\textwidth]{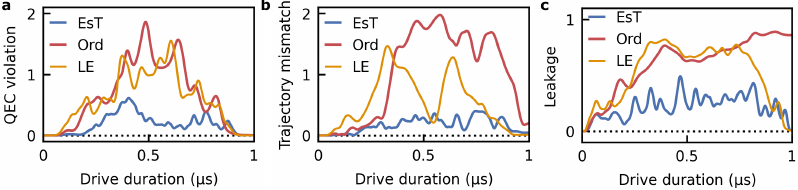}
\caption{\label{fig:app_combined_schematic}\textbf{EsT, LE and Ord gate comparison:} QEC violation, leakage and trajectory mismatch in instantaneous subspaces are calculated from lossless simulations for EsT, LE and Ord $X$ gate. 
The EsT and Ord gate metrics are the same from Fig.~\ref{fig:fig1}d-f, and are plotted again for ease of comparison with LE gate.}
\end{figure*}

In this section, we present additional results from numerical simulations that support the experimental data and provide more insight into the system dynamics. 
In Fig.~\ref{fig:app_combined_schematic}, we plot the QEC violation, leakage and trajectory mismatch conditions defined in Eq.~\eqref{eqn:Delta_qec},~\eqref{eqn:leakage_t},~\eqref{eqn:traj_mismatch} for the EsT, LE and Ord $X$ gates.
Leakage and trajectory mismatches are calculated for starting in code space state $|0_L\rangle$.
The plot shows that the violation of these metrics by the LE gate is lower than the Ord gate but higher than the EsT gate.
This similar hierarchy in fidelity for the three types of gates is observed in Fig.~\ref{fig:app_X_gate_expt}.
 
\begin{figure}[h!]
\includegraphics[scale = 0.95]{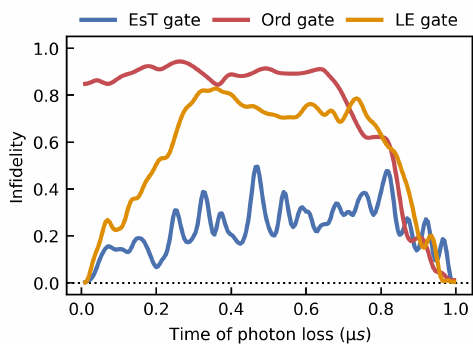}
\caption{\label{fig:app_err_in_time}\textbf{Gate fidelity for errors at a specific time:} We plot the infidelity in the error space at the end of the operation if a single photon loss happens at a specific time during the gate. 
Final infidelity is plotted as a function of this time of photon loss for the $X$ gate. 
We see that EsT gate has lowest infidelity owing to the ET and error space fidelity optimization. 
Ord gate performs worst and LE gate performance is in between Ord and EsT gates.
}
\end{figure}

In Fig.~\ref{fig:app_err_in_time}, we present another comparison of the three types of $X$ gate, by evaluating final state fidelity conditional on a photon loss error occurring at time $t$.
We compute this by evolving an initial state to time $t$ with Schrodinger equation, apply the photon loss operator and evolve the resulting state for the remaining pulse duration.
The final fidelity with respect to the target state in the error space is then averaged over the six cardinal state in code space as our initial states.
As expected we observe average infidelity follows the hierarchy of $\text{Ord}> \text{LE} > \text{EsT}$. 
The infidelity also decreases for errors occurring later in the pulse, since a larger fraction of the gate undergoes ideal evolution.

\begin{figure*}[t]
\includegraphics[width = 0.8\textwidth]{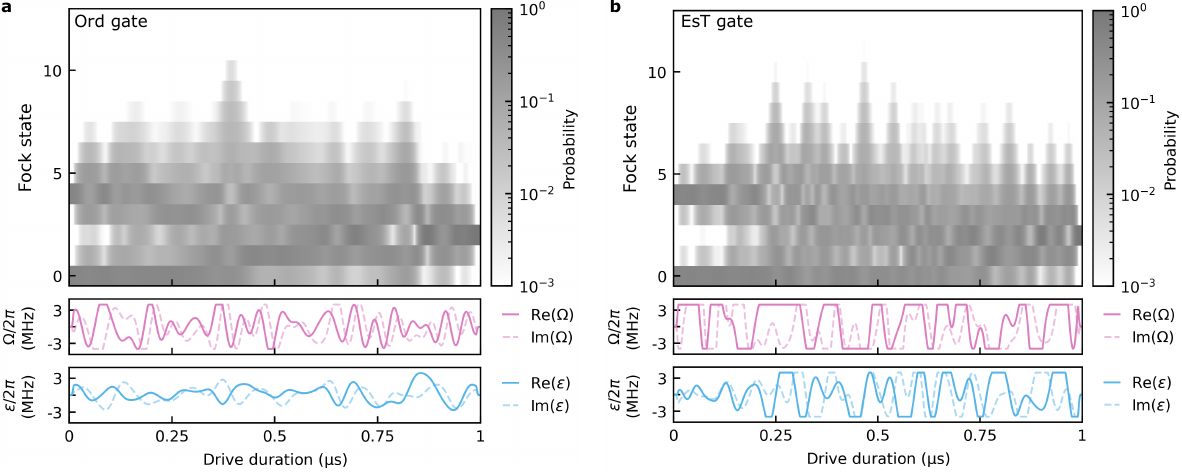}
\caption{\label{fig:app_dynamics}\textbf{Dynamics in the driven oscillator:} This plot shows the dynamics of the oscillator during $X$ gate on the binomial code and the optimized pulse shapes applied on the qubit and oscillator. 
(a) It shows the occupation of Fock states during the operation as color for Ord gate where gray means occupied and white means unoccupied. 
Bottom two panels show the complex amplitudes applied on qubit (top) and oscillator (bottom). 
(b) This plot also shows the oscillator Fock state occupation and complex drive amplitudes for the EsT gate. 
}
\end{figure*}

We also present the optimized pulse shapes and dynamics for the Ord and EsT gate in Fig.~\ref{fig:app_dynamics}a and ~\ref{fig:app_dynamics}b respectively.
The dynamics shows the probability of occupation of different Fock levels when starting from $|0_L\rangle$.
The pulse shape and dynamics during the EsT gate is more complicated than the Ord gate. 
But both gates achieve the same final state and same maximum active Fock state level with transient population greater than $0.01$.

\begin{figure*}[t]
\includegraphics[width = 0.85\textwidth]{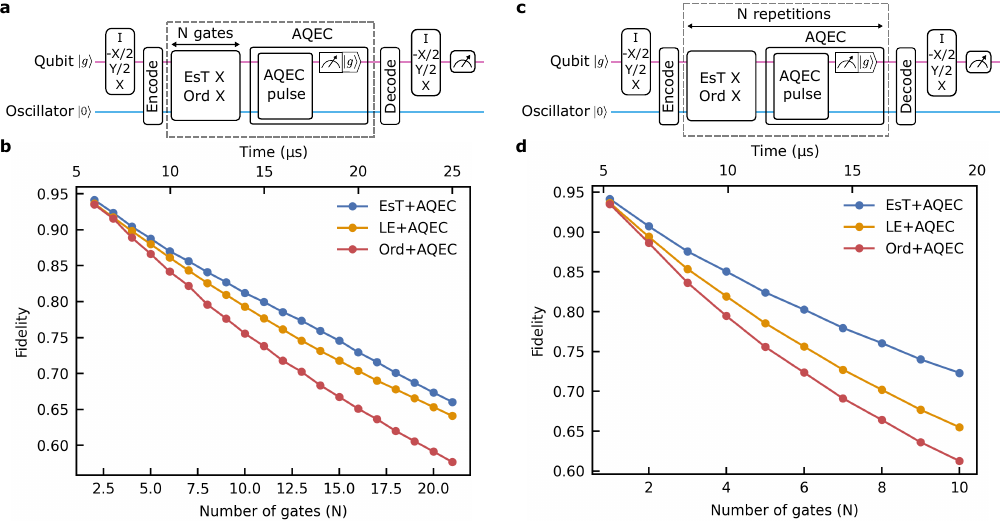}
\caption{\label{fig:app_aqec_sim}\textbf{Gate performance with AQEC:}(a) Pulse schematic used for simulation results in (b). 
We apply N gates before applying a single AQEC pulse. 
(b) This plot shows the performance scaling for EsT, LE, Ord gates where AQEC pulse is applied once after N gates. 
This plot uses coherence parameter and pulses used in our experiment. 
(c) Pulse schematic used for simulation results in (d). 
We apply an AQEC pulse after each gate. 
(d) This plot shows the performance scaling for EsT, LE, Ord gates where AQEC pulse is applied after each gate. 
This plot uses coherence parameter and pulses used in our experiment. 
These results are a proof of performance improvement of EsT gate without control errors. 
The dashed square boxes in the pulse schematic are the pulses that are simulated with losses for obtaining the fidelity decay plots.}
\end{figure*}

In Fig.~\ref{fig:app_aqec_sim}a we present the pulse schematic used for the experimental sequence shown in Fig.~\ref{fig:fig4}.
We simulate this sequence to obtain fidelity as a function of number of gates applied before AQEC operation to evaluate the performance in absence of control errors. 
The simulation results for three types of $X$ gate are shown in Fig.~\ref{fig:app_aqec_sim}c.
This validates our experimental observation the EsT gate significantly outperforms the Ord gate. 
Further, the LE gate yields an intermediate performance, outperforming Ord gate but not performing as good as the EsT gate.

Fig.~\ref{fig:app_aqec_sim}b shows a pulse sequence where an AQEC operation is applied after each gate.
The simulation results for three gates, using experimental parameters, are shown in Fig.~\ref{fig:app_aqec_sim}d. 
The absolute fidelity decays rapidly, primarily due to the infidelities of the AQEC pulse itself. 
However, the relative per-gate fidelity improvement for the EsT gate is higher than the Ord gate. 

Finally, we isolate the sources of error in our gate operations using numerical simulation and models fitted to experimental data. 
The $X$ and $H$ gate fidelities for various scenarios are summarized in Table.~\ref{tab:fidelity_comparison}. 
We observe excellent agreement between simulated and experimentally extracted fidelities.
The slight mismatch primarily arises from control errors and Hamiltonian parameter fluctuations.
Below we point a few key observations from the table. 
Fidelities calculated without qubit errors are similar to those without oscillator errors, indicating that both loss channels are significant limits to gate performance.
There is a clear enhancement in ET fidelity, validating our claim that error-transparency optimization substantially improves overall gate performance. 

\begin{table*}[t]
    \centering
    \renewcommand{\arraystretch}{1.2} 
    \setlength{\tabcolsep}{5pt}
    \begin{tabular}{|l|c|c|c|c|c|c||c|c|c|c|} 
        \hline
        & \multicolumn{6}{c||}{\textbf{X Gate}} & \multicolumn{4}{c|}{\textbf{H Gate}} \\
        \cline{2-11}
        & \multicolumn{3}{c|}{\textbf{Simulation}} & \multicolumn{3}{c||}{\textbf{Experiment}} & \multicolumn{2}{c|}{\textbf{Simulation}} & \multicolumn{2}{c|}{\textbf{Experiment}} \\
        \cline{2-11}
        \textbf{Metric} & \textbf{EsT} & \textbf{LE} & \textbf{Ord} & \textbf{EsT} & \textbf{LE} & \textbf{Ord} & \textbf{EsT} & \textbf{Ord} & \textbf{EsT} & \textbf{Ord} \\
        \hline \hline
        \multicolumn{11}{|l|}{\textbf{With all errors}} \\
        \hline
        Net fidelity          & 0.974 & 0.974 & 0.975 & -- & -- & -- & 0.975 & 0.977 & -- & -- \\
        \hline
        Code space fidelity   & 0.990 & 0.988 & 0.987 & 0.985 & 0.971 & 0.984 & 0.991 & 0.989 & 0.988 & 0.986 \\
        \hline
        Error space fidelity  & 0.992 & 0.990 & 0.190 & 0.960 & 0.930 & 0.200 & 0.992 & 0.218 & 0.967 & 0.200 \\
        \hline
        ET fidelity           & 0.676 & 0.476 & 0.430 & 0.750  & 0.400 & 0.350 & 0.669 & 0.347 & 0.750  & 0.350 \\
        \hline
        Ideal QEC             & 0.983 & 0.980 & 0.978 & -- & -- & -- & 0.985 & 0.981 & -- & -- \\
        \hline
        \multicolumn{11}{|l|}{\textbf{No qubit errors}} \\
        \hline
        Net fidelity          & 0.985 & 0.986 & 0.984 & -- & -- & -- & 0.987 & 0.987 & -- & -- \\
        \hline
        Code space fidelity   & 0.996 & 0.995 & 0.992 & 0.985 & 0.980 & 0.983 & 0.997 & 0.995 & 0.986 & 0.984 \\
        \hline
        Error space fidelity  & 0.997 & 0.996 & 0.350 & 0.960  & 0.960 & 0.400 & 0.997 & 0.350 & 0.969 & 0.350 \\
        \hline
        ET fidelity           & 0.843 & 0.638 & 0.533 & 0.850  & 0.500 & 0.500 & 0.898 & 0.381 & 0.880  & 0.400 \\
        \hline
        Ideal QEC             & 0.994 & 0.991 & 0.986 & -- & -- & -- & 0.996 & 0.989 & -- & -- \\
        \hline
        \multicolumn{11}{|l|}{\textbf{No oscillator errors}} \\
        \hline
        Net Fidelity          & 0.985 & 0.987 & 0.986 & -- & -- & -- & 0.987 & 0.989 & -- & -- \\
        \hline
    \end{tabular}
    \caption{\textbf{Fidelity comparison of $X$ and $H$ gates in experiment and simulation:} The fidelity values noted for simulations are obtained from propagating the states with master equation solver for the gate Hamiltonian with all device parameters. It does not account for any state preparation or measurement loss. Values in the experiment sections are extracted from models that fit the experimental data.}
    \label{tab:fidelity_comparison}
\end{table*}

\section{Additional experimental data\label{app:add_expt}}

\begin{figure*}[t]
\includegraphics[width = \textwidth]{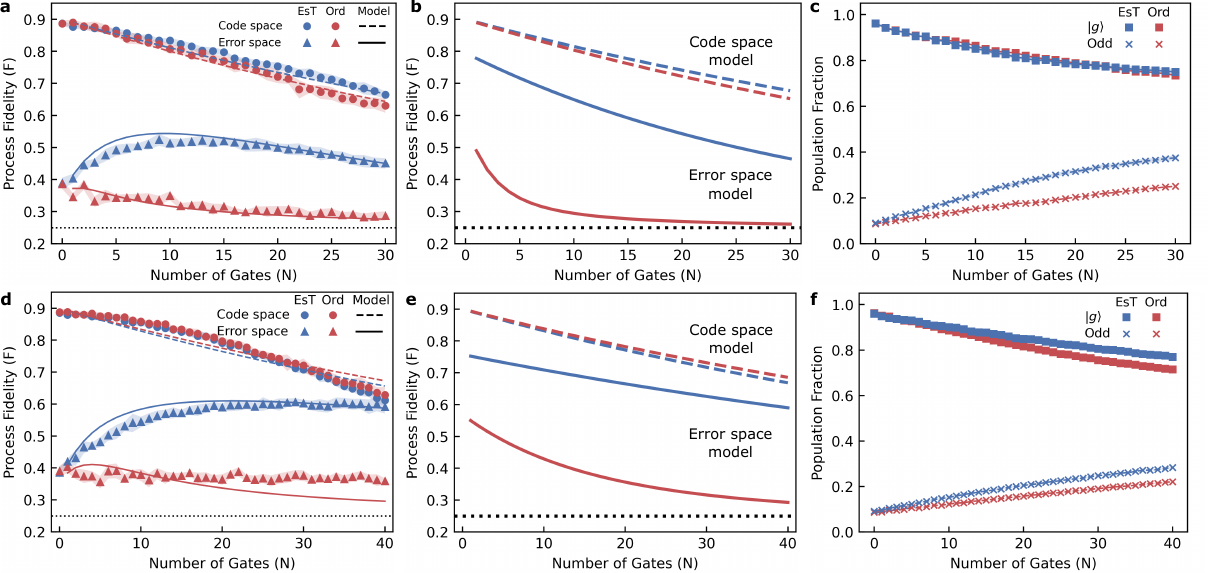}
\caption{\label{fig:app_HT_gate_expt}\textbf{Code and error space process fidelity:} (a) Experimentally measured code and error space process fidelities with number of gates applied for $H$ gate with modeled decay curves shown. (b) We plot the decay curves without parity selection error used for selecting code and error space states. 
(c) experimentally measured probability of finding qubit in ground state $|g\rangle$ and measuring a state in odd parity at the end of the gate operation for H gate. Similarly for T gates, (d) experimentally measured process fidelities and modeled decay curves, (e) decay curves without parity selection errors, and (f) experimentally measured probabilities finding qubit in ground state $|g\rangle$ and measuring a state in odd parity at the end of the gate operation are plotted.
This data set and models has been used to extract the per-gate fidelity shown in Fig.~\ref{fig:fig3}(d).
}
\end{figure*}

\begin{figure*}[t]
\includegraphics[width = \textwidth]{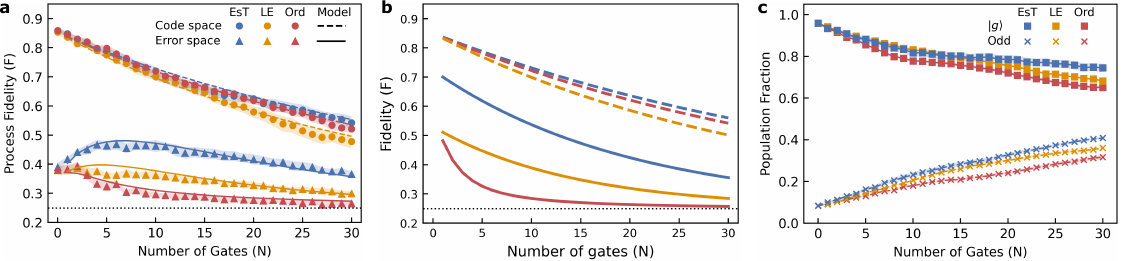}
\caption{\label{fig:app_X_gate_expt}\textbf{LE gate performance for X operation:} (a) Experimentally measured code and error space performance of LE gate, optimized for code and error space process fidelity but not ET, plotted alongside EsT and Ord gate experiments in Fig.~\ref{fig:fig3}(b). 
Modeled decay curves are shown as lines. 
The LE gate performance is similar to other gates in code space.
In error space LE gate performs better than Ord gate as the fidelity of operation in error space is higher but it performs worse than EsT gates as it is not optimized for the ET condition.
(b) Model decay curves without parity selection errors are plotted.
(c) experimentally measured probabilities finding qubit in ground state $|g\rangle$ and measuring a state in odd parity at the end of the gate operation are plotted for all three gates.
}
\end{figure*}

\begin{figure*}[t]
\includegraphics[width = \textwidth]{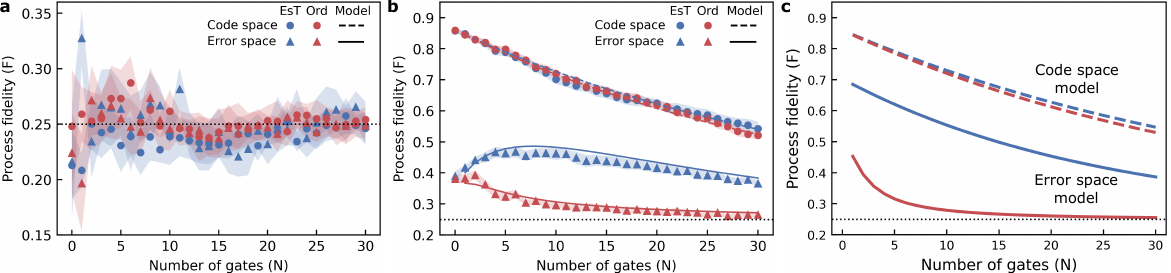}
\caption{\label{fig:app_qb_e_expt}\textbf{Code and error space process fidelity with qubit errors:} (a) code and error space process fidelities measured for $X$ gate post-selected on the cases where qubit is found in the excited state $|e\rangle$ at the end of gate operations. 
It shows that a qubit error scrambles the quantum information completely.
(b) We combine the cases of qubit in ground state or qubit in excited state after the gates to plot process fidelities unselective of qubit errors. 
The plot still shows a significant improvement of error space process fidelity for EsT gate compared to Ord gate although the absolute fidelity numbers are lower.
(c) Modeled decay curve without parity selection error further validates the performance improvement of EsT gate.
}
\end{figure*}

\begin{figure*}[t]
\includegraphics[width = 0.85\textwidth]{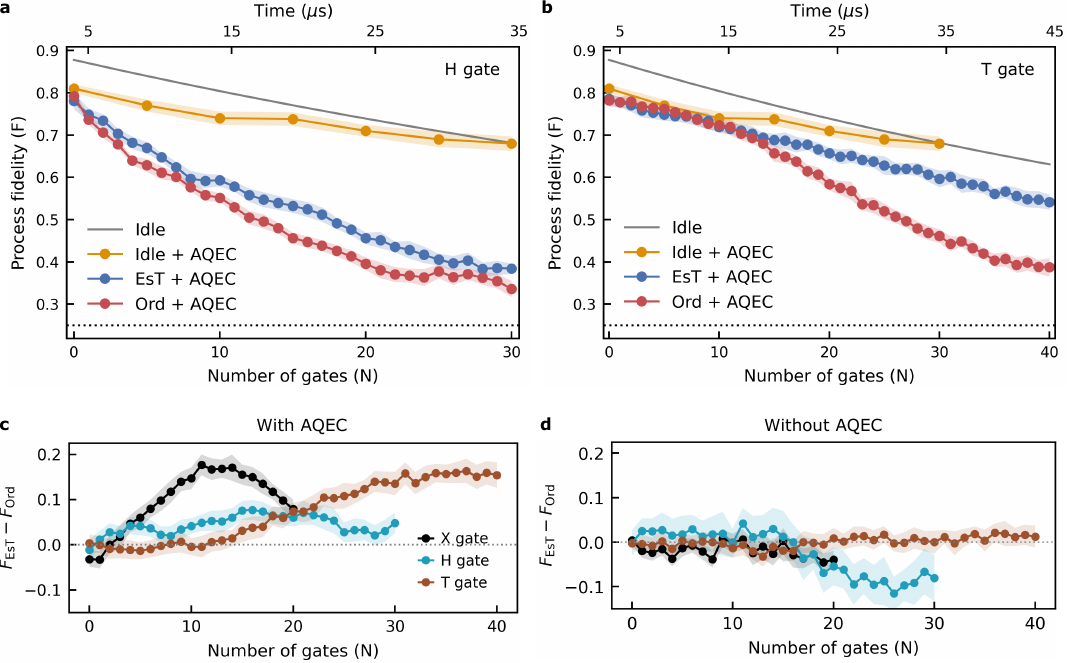}
\caption{\label{fig:app_aqec_expt}\textbf{Gate performance improvement with AQEC:} Process fidelity after applying a single AQEC pulse after applying $N$ gates for EsT and Ord (a) $H$ gate and (b) $T$ gate. 
Process fidelity decay for idling and for applying an AQEC pulse after idling for an equivalent duration of N gates. 
The horizontal dashed line at process fidelity of 0.25 indicates a completely incoherent process and shaded regions around the data points signify 1 standard deviation of fidelities for repeated measure-ments.
(c) Performance improvement comparison of different gates as a function of number of gates applied before AQEC pulse. 
This is the same plot shown in Fig.~\ref{fig:fig4}(c).
(d) Performance comparison of EsT and Ord gates with number of gates without AQEC. 
This shows that the EsT gates performed worse than Ord gate due to Hamiltonian parameter fluctuations. 
Nevertheless, after AQEC the EsT gate performance improves compared to Ord gates.
}
\end{figure*}

In this section, we present extended dataset that was used for main text figures and additional experimental data that supports our claims.

For calculating the process fidelity shown in Fig.~\ref{fig:fig3}-\ref{fig:fig5}, we first perform process tomography~\cite{NielsenChuang2010, PRXQuantum.6.030202}. 
For this we prepare the qubit in four cardinal points of the Bloch sphere: $\{ |g\rangle, |e\rangle, (|g\rangle + |e\rangle)/\sqrt(2), (|g\rangle + i|e\rangle)/\sqrt(2) \}$.
Then the information is encoded onto the logical Bloch sphere with the encode operation and our process (gates, QEC, parity measurement as applicable in the respective experiment) is performed. 
Finally, the information is decoded back to the ancilla with decode operation and the ancilla state is measured in all three Pauli basis. 
From this we obtain the $4\times 4$ process matrix $\chi_G$. 
The fidelity is defined as the overlap of the process matrix with the target process matrix $\chi_{\text{target}}$ as $F = \text{tr}(\chi_G \chi_{\text{target}})$. 
The encode-decode process (which should be an identity operation) has a process fidelity of $\sim91\%$.

Fig.~\ref{fig:app_HT_gate_expt} shows additional experimental data for code and error space process fidelity for $H$ and $T$ gate. 
Experimentally measured data is plotted as circles (for code space) and triangle (for error space). 
Solid lines for error space and dashed lines for code space are models that fit the experimental data. 
Shaded regions represent $1$ standard deviation of fidelities measured in repeated experiments. 
Fig.~\ref{fig:app_HT_gate_expt}a and Fig.~\ref{fig:app_HT_gate_expt}b respectively shows the experimental data and model without parity selection errors for $H$ gate.
These two plots confirm similar code space performance for EsT and Ord gate but significant improvement in performance for EsT gate in error space over the Ord gate. 
Fig.~\ref{fig:app_HT_gate_expt}c shows the experimentally measured probabilities of qubit being in ground state at the end of the gate operation (square points) and the probability that the state is an odd parity state (crosses) i.e. in error space. 
The state is found in odd parity state if a photon loss error happens during the operation. 
This plot shows error rates for qubit relaxation and oscillator photon loss are comparable for the H gate. 
However, the ET gate has slightly higher probability of being in the odd state at the end of the operation than the Ord gate.
This can be attributed to higher average photon number for the ET gate during the operation.

Similarly, Fig.~\ref{fig:app_HT_gate_expt}d and Fig.~\ref{fig:app_HT_gate_expt}e respectively shows the experimental data and model without parity selection errors for $T$ gate.
They also confirm the similar performance in code space and different performance in error space for the $T$ gate. 
Fig.~\ref{fig:app_HT_gate_expt}f shows that probabilities of qubit error and oscillator photon loss error are comparable for both ET and Ord $T$ gate. 
We do not fully understand the reason for the small mismatch of error space model and experimental data for the Ord $T$ gate.  

We then turn to experimentally verifying the performance of LE gate described in App.~\ref{fig:app_X_gate_expt}. 
This gate is optimized for maximizing fidelity in code and error space but not for error transparency. 
The code and error space performance is similarly plotted in Fig.~\ref{fig:app_X_gate_expt}a alongside EsT and Ord gate data. 
The data for EsT and Ord gate is the same as in Fig.~\ref{fig:fig3}b, and is presented here for ease of performance comparison. 
Fig.~\ref{fig:app_X_gate_expt}b shows the modeled gate fidelities without parity selection errors. 
This data confirms that LE gate performance is better than Ord gate as the operation is high fidelity in error space but is worse than EsT gate as it is not optimized for error transparency. 
Hence a photon loss during the operation still causes a significant loss in fidelity of the operation. 
It signifies the importance of the error transparency cost defined in Eq.~\eqref{eqn:app_et_cost} for optimization. 
Similar to the previous figure, experimentally measured probabilities of qubit being in ground state and the probability of finding an odd parity state (an error space state) is shown in Fig.~\ref{fig:app_X_gate_expt}c.
The error rates due to qubit relaxation and oscillator photon loss are comparable and remain consistent for all three Ord, LE and EsT $X$ gates. 

The code and error space process fidelity data presented so far is post-selected on qubit being in ground state to avoid complications of qubit errors and those propagating into parity selection and decode operations. 
However, qubit errors are an integral part of these gate operations. 
So it is important to understand the gate dynamics when a qubit relaxation error happens during the gate.
We can similarly perform code and error space process fidelity experiments unselective of qubit errors. 
Fig.~\ref{fig:app_qb_e_expt}a shows the experimental data for EsT and Ord X gate performance for the cases when qubit is found in the excited state at the end of the operation. 
If these cases are considered, the absolute values of code and error space process fidelities will drop.
The final fidelities for the EsT and Ord gate unselective of qubit errors is presented in Fig.~\ref{fig:app_qb_e_expt}b, and the modeled decay curves without parity selection errors are shown in Fig.~\ref{fig:app_qb_e_expt}c.
This data shows that without selecting for qubit errors the absolute numbers for code and error space process fidelities are lower.
However, the similarity in performance in code space and improvement in performance in the error space for EsT gate remains true.

Finally we present additional data for AQEC sequence shown in Fig.~\ref{fig:fig4}a for $H$ and $T$ gate respectively in Fig.~\ref{fig:app_aqec_expt}a and Fig.~\ref{fig:app_aqec_expt}b. 
This data was used to plot the performance improvement comparison of EsT and Ord gates for the three $X$, $H$ and $T$ operations shown in Fig.~\ref{fig:fig4}c. 
The same plot is shown in Fig.~\ref{fig:app_aqec_expt}c for ease of comparison with Fig.~\ref{fig:app_aqec_expt}d which shows the performance difference of EsT and Ord gate without the AQEC pulse. 
As we can see the EsT gate performance is either worse or similar to the performance of Ord gate.
This is primarily due to parameter fluctuations causing the EsT gate fidelity to be worse than Ord gate.
However, despite worse performance of EsT gate without AQEC, the performance improvement with the AQEC pulse shows that EsT gate is preserving target state amplitude and phase coherence in the error space.
In future, including robustness to parameter fluctuation as an optimization objective and doing hardware feedback based calibration can be used for removing such errors. 

It is also worth noting that, while $H$ gate shows the maximum performance improvement in error space fidelity for the EsT gate as shown in Fig.~\ref{fig:fig3}d, it also has a higher probability of photon loss error during the operation as indicated by the probability of measuring odd parity state shown in Fig.~\ref{fig:app_HT_gate_expt}c. 
Our AQEC operation is also not very high fidelity. 
Hence the combination of higher fraction of states in the odd space, low AQEC fidelity and a low fidelity of the gate in absence of AQEC due to control errors causes the $H$ gate performance with AQEC to not be significantly higher compared to Ord gate as presented in Fig.~\ref{fig:fig4}c and Fig.~\ref{fig:app_aqec_expt}c. 

 \section{Modeling code and error space fidelity\label{app:espace_model}
}
For the binomial `kitten' code, code space states are even parity and error space states are odd parity. 
Therefore, measurements of Fock state parity, allows categorization of states as being potentially in the code or error spaces.
This parity measurement has infidelity due to qubit relaxation or dephasing during the measurement, readout assignment error, and finite pulse bandwidth of the qubit $\pi/2$ pulses used for parity mapping and control errors. 
Besides this, due to highly asymmetric population in even and odd states for the binomial code, the sparse error space population means that the measurement infidelity contributes a particularly high proportion of the error in parity selection. 
We illustrate and analyze this issue in this subsection.

Before the parity measurement, probability that the state in the cavity is odd is written as $P(O)$. 
This depends on the intrinsic error rate of the oscillator, duration of the gate operation, mean photon number in the oscillator and importantly the number of gates applied.
Parity mis-assignment error is termed as $\epsilon_{\mathrm{parity}}$ which signifies the probability that parity selection maps to wrong parity state.
This is assumed to be symmetric for odd and even states. 
For the odd parity state, we can write the probability that a truly odd state is measured to be odd ($M_O$) as  
\begin{equation}
    P(M_O|O) = 1 - \epsilon_{\mathrm{parity}}
\end{equation}
So the probability that a state measured in odd is actually odd is given by 
\begin{equation}
    P(O|M_O) = P(M_O|O) P(O) /P(M_O)~, \\
\end{equation}
where, $P(M_O) = P(M_O|O) P(O) + P(M_O|E)P(E) = P(M_O|O) P(O) + (1 -P(M_O|O))P(E) $. 
Here, $P(E) = 1 - P(O)$ is the probability of the state being even before the parity measurement and $P(M_O|E) = 1 - P(M_O|O)$ is the probability that an even parity state is measured to be odd parity.
For the `true' odd states passing through the selection, we label with fidelity $F_{\mathrm{true}}$. 
For the even states selected as odd, we label with fidelity $F_{\mathrm{alias}}$. 
So the measured fidelity will be, 
\begin{equation}
    F_{\mathrm{measured}}= P(O|M_O) F_{\mathrm{true}} + ( 1 - P(O|M_O)) F_{\mathrm{alias}}~.
\end{equation}
The decode operation used in our experiment is designed to map odd states from the error space Bloch sphere to the qubit Bloch sphere. 
So when this decode operation is applied on the even states no relevant quantum information is mapped to the qubit Bloch sphere. 
So $F_{\text{alias}}$ is taken to be $0.25$ in our model.

In the $F_{\mathrm{true}}$, we also have two components. 
A fraction of odd states are generated due to photon loss during the gate operation. 
We define this probability as $P(\mathrm{O, gate})$ and the corresponding fidelity as $F_{\mathrm{O, gate}}$. 
The other fraction of odd states were odd because of states that were odd at the start of the gate because of error during the encode process. 
This is essentially a state preparation error. 
We define this probability as $P(\mathrm{O, prep})$ and fidelity as $F_{\mathrm{O, prep}}$. 
The readout error and loss during the parity measurement can be thought of as a parity mapping error and can be absorbed into $\epsilon_{\mathrm{parity}}$. 
So, we can write
\begin{equation}
    F_{\mathrm{true}} = P(\mathrm{O, gate}) F_{\mathrm{O, gate}} + P(\mathrm{O, prep}) F_{\mathrm{O, prep}}~.
\end{equation}
We are interested in the quantity $F_{\mathrm{O, gate}}$.
Here, $P_{\mathrm{O, gate}}$ and  $P(\mathrm{O, prep})$ are normalized such that they add up to 1. 
If not normalized they should add up to $P(O)$. 

A similar analysis can be done for the even states. 
However, it is worth noting that in our case of binomial `kitten' code, when applying logical operations starting from code space (even parity states), we will most often end in even parity states.
Only when a single photon loss error happens during the gate operation, we will end up in error space (odd parity states). 
Hence, this parity selection error massively affects the fidelity of states measured in the error space whereas fidelity of the states measured in code space is much closer to the true even parity population. 

\begin{figure}[h]
\includegraphics[scale = 0.9]{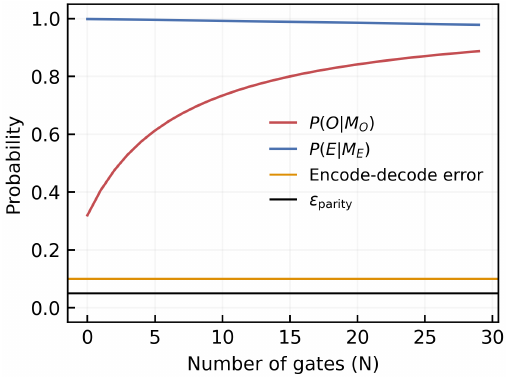}
\caption{\label{fig:app_err_model}\textbf{Error model for parity selection errors:} We plot the error probability of errors in parity measurement ($\epsilon_{\text{parity}}$) and encode-decode process. 
We also plot the probability that a state is truly odd if it is measured odd ($P(O|M_O)$) and that a state is truly even if it is measured even ($P(E|M_E)$).
This summarizes the contribution of various error channels in code and error space process fidelity measurement. 
}
\end{figure}

We show the probabilities of different errors in Fig.~\ref{fig:app_err_model}.
$\epsilon_{\mathrm{parity}}$ is fixed at $0.05$ (black horizontal line) for our experiment which is fixed for any number of gates applied. 
The error in the encode-decode process for preparing a logical state and reading it is about $9\%$ is also shown as orange horizontal line.
The plot shows $P(O|M_O)$ and $P(E|M_E)$ with number of applied gates. 
The numbers used in this model follow the device parameters for our experiment. 
The $P(O|M_O)$ as a function of number of gates show that the probability that a state measured odd is truly odd increases with number of gates as more number of gates mean longer duration of operation and hence more photon loss errors in the oscillator. 
And the large change in this probability is the reason for large difference in decay profile, between the experimentally measured fidelity and the parity-selection error free fidelities as seen in Fig.~\ref{fig:fig3}, \ref{fig:app_HT_gate_expt}, \ref{fig:app_X_gate_expt}, \ref{fig:app_qb_e_expt}.

Next, we model the true code and error space process fidelities.
For the code space, we model with an exponential fidelity decay due to coherence times and control errors.
These can be combined to write the fidelity with number of gates as 
\begin{equation}
    F_{\mathrm{code}}(N) = A \exp{\left(-\gamma_C N \right)} +B~,
\end{equation}
where, $A$ and $B$ are constants and $\gamma_C$ is the error rate per gate in the code space.
We set $B = 0.25$ as the point where the process is completely incoherent and choose $A$ accordingly to match the process fidelity of encode-decode process measured in experiments.
$\gamma_C$ depends on the gate duration, coherence times of the cavity and qubit and control errors.

For the error space, we need a more complicated model as the fidelity measured at $N$-th gate includes photon loss at the $k$-th step for $k\leq N$ and error space gate fidelity for the $(N-k)$ gates.
The photon loss is equally likely during each gate operation.
For the $N$-th gate process fidelity measured in error space, the photon loss happened in between $k=1$ to $N$ gates with probability of loss $1/N$ per gate. 
if the photon loss happened at $k$-th step, then fidelity due to this branch will be
\begin{equation}
    f_k =F_C^{k-1} F_{\mathrm{err|jump}} F_E^{N-k}~,
\end{equation}
where, $F_C = e^{-\gamma_C}$ is the fidelity per gate in code space and $F_E = e^{-\gamma_E}$ is the fidelity per gate in error space and $F_{\mathrm{err|jump}}$ is the fidelity when there is a jump from code to error space during the gate operation. $\gamma_E$ is the error rate in error subspace which depends on the gate duration, coherence times of the cavity and qubit and control errors for the error space operation.
$F_{\mathrm{err|jump}}$ is related to the error transparency of our pulse: the more error transparent the pulse is, the higher will be $F_{\mathrm{err|jump}}$.
So the total fidelity in the error space for the $N$-th gate should be, $F_{\mathrm{error}}(N) = D.\frac{1}{N} \sum_{k=1}^Nf_k$.
D is the scaling that includes encode decode process fidelity.
This is a geometric series which can be summed to
\begin{equation}
    F_{\mathrm{error}}(N) = \frac{D F_{\mathrm{err|jump}}}{N} \frac{F_C^N - F_E^N}{F_C - F_E}  + G~,
    \label{eqn:espace_model}
\end{equation}
where, the constant $G=0.25$ is the lower limit of fidelity and signifies a completely incoherent process.
$D$ is then chosen to match the encode-decode process fidelity measured in experiment in the error subspace.
We choose $D=0.6$ for our model.

This model works well for the Ord gate where $F_C$ and $F_E$ are quite different.
But for ET gate, $F_C \approx F_E$ and we need to take the limit of this Eq.~\eqref{eqn:espace_model}.
Taking the limit $F_C - F_E \rightarrow 0$, for the ET gate, we have
\begin{equation}
    F_{\mathrm{error, ET}}(N) = D \cdot F_{\mathrm{err|jump}}  \cdot F_C^{N-1} + G~.
\end{equation}

We have not considered double photon loss errors but they remain smaller than $4\%$ at the \SI{30}{\micro\second} timescale we have considered in our experiments. 
So our first order model fits the experimental data reasonably well (Fig.~\ref{fig:fig3}, \ref{fig:app_HT_gate_expt}, \ref{fig:app_X_gate_expt}, \ref{fig:app_qb_e_expt}) and allows us to extract relevant error rates and fidelities.

\newpage
\clearpage
\bibliography{sr_bibliography}

\end{document}